\documentclass[journal]{IEEEtran}

\usepackage{booktabs}
\usepackage{array}
\usepackage{longtable}
\usepackage{float}
\usepackage{lipsum} 
\usepackage{cite}
\usepackage{amsmath,amssymb,amsfonts,bm}
\usepackage{algpseudocode}
\usepackage{algorithm}
\usepackage{listings}
\usepackage{tabularx}
\usepackage{graphicx}
\usepackage{textcomp}
\usepackage[dvipsnames]{xcolor}
\usepackage[caption=false]{subfig}
\usepackage{glossaries}
\usepackage{hyperref}
\usepackage{multirow}
\usepackage{comment}
\usepackage{amsmath}
\usepackage{cuted}
\usepackage{adjustbox}
\usepackage{svg}
\usepackage{cases}
\usepackage{amsthm}

\usepackage{acronym}
\usepackage{pict2e}

\usepackage{todonotes}

\def\BibTeX{{\rm B\kern-.05em{\sc i\kern-.025em b}\kern-.08em
    T\kern-.1667em\lower.7ex\hbox{E}\kern-.125emX}}

\newsavebox{\ORCIDlogo}
\savebox{\ORCIDlogo}{%
\setlength{\unitlength}{\dimexpr 1em/256\relax}%
\begin{picture}(256,256)%
  \color[HTML]{A6CE39}\put(128,128){\circle*{256}}%
  \color{white}%
  \put(78.6,199.2){\circle*{20}}%
  \moveto(70.9,176,9)\lineto(86.3,176,9)\lineto(86.3,69.8)\lineto(70.9,69.8)%
  \closepath\fillpath%
  \moveto(108.9,176.9)\lineto(150.5,176.9)%
  \curveto(190.1,176.9)(207.5,148.6)(207.5 ,123.3)%
  \curveto(207.5,95,8)(186,69.7)(150.7,69.7)%
  \lineto(108.9,69.7)%
  \closepath\fillpath%
  \color[HTML]{A6CE39}%
  \moveto(124.3,83.6)\lineto(148.8,83.6)%
  \curveto(183.7,83.6)(191.7,110.1)(191.7,123.3)%
  \curveto(191.7,144.8)(178,163)(148,163)%
  \lineto(124.3,163)%
  \closepath\fillpath%
\end{picture}%
}

\newcommand\orcidicon[1]{\href{https://orcid.org/#1}{\usebox{\ORCIDlogo}}}    
    
\acrodef{ADS-B}{Automatic Dependent Surveillance-Broadcast}
\acrodef{CFO}{Carrier Frequency Offset}
\acrodef{CNN}{Convolutional Neural Network}
\acrodef{CSI}{Channel State Information}
\acrodef{DL}{Deep Learning}
\acrodef{FPGA}{Field-Programmable Gate Array}
\acrodef{IoT}{Internet of Things}
\acrodef{IQ}{In-Phase and Quadrature}
\acrodef{LoRa}{Long Range}
\acrodef{LSTM}{Long Short-Term Memory}
\acrodef{ML}{Machine Learning}
\acrodef{NN}{Neural Network}
\acrodef{PCA}{Principal Component Analysis}
\acrodef{PHY}{Physical}
\acrodef{RF}{Radio Frequency}
\acrodef{RFF}{Radio Frequency Fingerprinting}
\acrodef{RFMLS}{Radio Frequency Machine Learning System}
\acrodef{SDR}{Software-Defined Radio}
\acrodef{SEI}{Specific Emitter Identification}
\acrodef{SNR}{Signal-to-Noise Ratio}
\acrodef{SVM}{Support Vector Machine}
\acrodef{UAV}{Unmanned Aerial Vehicle}
\acrodef{UMOP}{Unintentional Modulation on Pulse}
\acrodef{PLC}{Power Line Communication}

\begin{document}

\title{The Chronicles of Radio Frequency Fingerprinting}

\author{Abdul Aziz \ {\orcidicon{0000-0002-3357-2158}} \and, Ingrid Huso \ {\orcidicon{0000-0002-5134-5792}} ~\IEEEmembership{Member,~IEEE} \and, \\ Savio Sciancalepore\ \orcidicon{0000-0003-0974-3639} ~\IEEEmembership{Member,~IEEE} \and, and Gabriele Oligeri\ \orcidicon{0000-0002-9637-0430} ~\IEEEmembership{Member,~IEEE}\and
\thanks{Abdul Aziz, Ingrid Huso, and Gabriele Oligeri are with the College of Science and Engineering, Hamad Bin Khalifa University, Doha, Qatar; \textit{\{abaz89721, ihuso, goligeri\}@hbku.edu.qa}. Savio Sciancalepore is with Eindhoven University of Technology, Eindhoven, Netherlands;\\ \textit{s.sciancalepore@tue.nl}.\\ This work was partially supported by the NATO SPS G8414 project.\\ (\emph{Corresponding author: Gabriele Oligeri, e-mail: goligeri@hbku.edu.qa.})}
}

\maketitle

\begin{abstract}
Radio Frequency Fingerprinting (RFF) has evolved from an early idea for radar emitter identification into a broad research field for wireless device identification and spectrum monitoring for security. Rather than presenting a conventional literature survey, this work provides a critical historical analysis of RFF organized around the field's major conceptual paradigm shifts from 1993 to 2026. We discuss the evolution of RFF across its fundamental methodological phases, beginning with early transient-based approaches, in which transmitter turn-on behavior, unintentional modulation, and hardware nonlinearities were treated as the primary fingerprint sources. We then examine the transition to digital communications, during which attention shifted to steady-state impairments and to engineered features extracted from signals. Next, we discuss the Machine Learning period, which standardized the RFF workflow around feature extraction, dimensionality reduction, and supervised classification, followed by the Deep Learning period, in which representation learning from raw IQ samples significantly improved performance and expanded the application space. Beyond a chronological list of methods and best practices, this paper critically examines the changing assumptions and persistent limitations that have driven these transitions. We highlight the central challenges that continue to shape the field, including channel dependence, receiver sensitivity, limited dataset realism, poor cross-domain generalization, open-set recognition, and adversarial robustness. We show that the current direction of RFF research is increasingly defined not by accuracy alone, but also by the need for credible, reproducible, and deployment-ready evaluation. By organizing more than three decades of work into a coherent narrative, this paper clarifies the evolution of RFF, identifies persistent limitations, and outlines the key research directions required to move the field toward dependable real-world adoption.
\end{abstract}

\maketitle

\section{Introduction}
\label{sec:introduction}

\begin{figure*}[t]
 \centering
 \includegraphics[width=\textwidth, angle = 0,trim = 0mm 30mm 0mm 10mm]{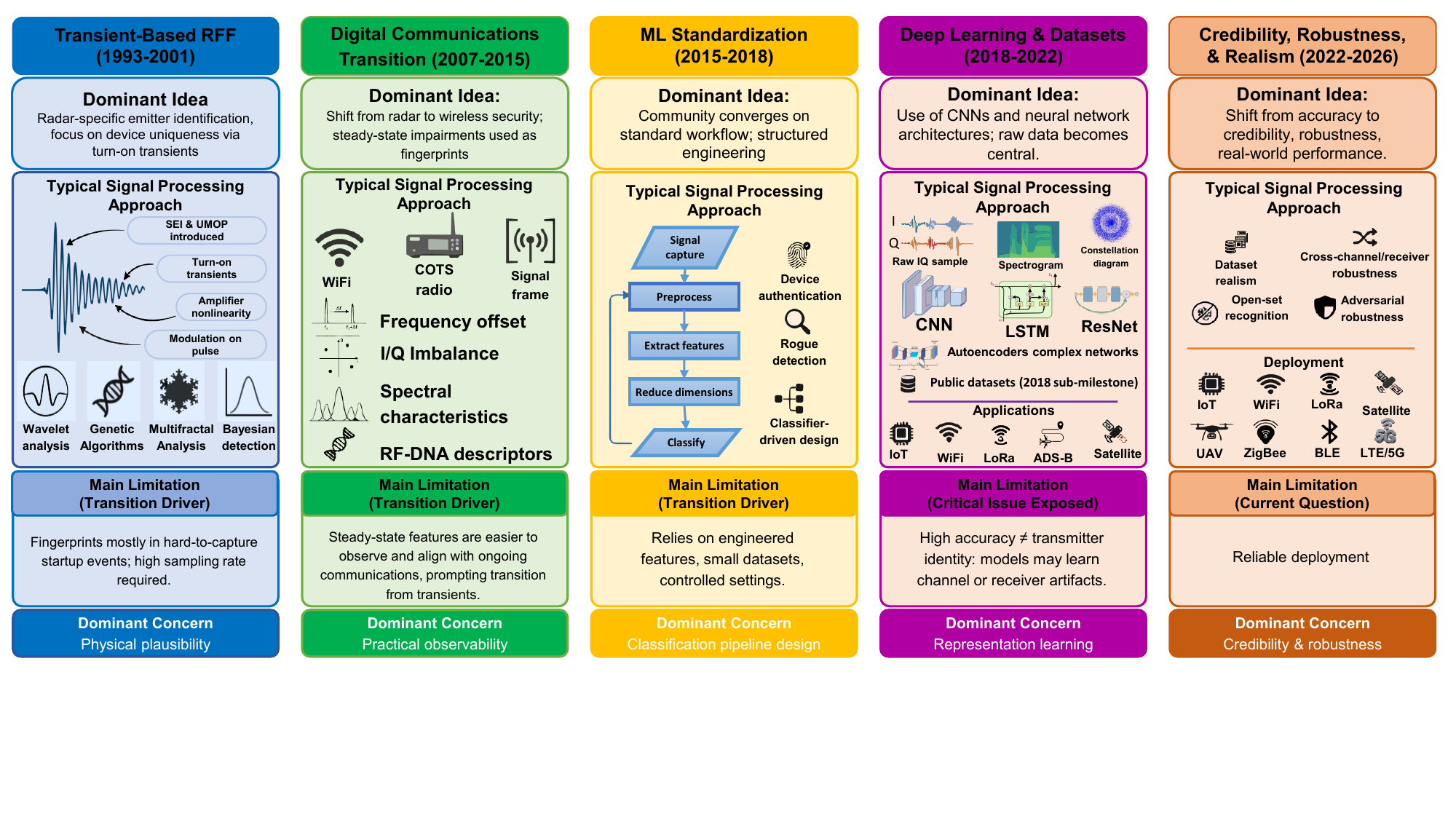}
 \caption{Three decades of Radio Frequency Fingerprinting research.}
 \label{fig:timeline}
\end{figure*}

\ac{RFF} has emerged recently as an effective approach to identify wireless transmitters by exploiting subtle hardware-dependent imperfections embedded in their emitted signals~\cite{jagannath_comprehensive_2022}. Unlike software credentials or higher-layer identifiers, these \ac{PHY}-layer features are difficult to replicate, making them a promising basis for device identification, authentication, and wireless security. Over the last three decades, \ac{RFF} has evolved from a niche idea rooted in radar-specific emitter identification into a broad research area spanning WiFi, \ac{IoT}, \ac{LoRa}, \ac{ADS-B}, satellite communications, and other wireless domains~\cite{Zhang2025}. As wireless systems have become more heterogeneous and security-critical, distinguishing devices based on \ac{PHY} signatures has become increasingly important.  

The evolution of \ac{RFF} has been closely tied to changing perspectives on what constitutes a meaningful and reliable fingerprint. Early work focused primarily on transient behavior during transmitter turn-on, motivated by the radar setting of the earliest work~\cite{toonstra_radio_1996_2}. These approaches employed dedicated signal processing techniques to extract fast-signal features associated with hardware-specific characteristics. While these methods established the physical plausibility of transmitter identification, they were also constrained by the difficulty of reliably capturing transients and the need for high sampling rates~\cite{ureten_wireless_2007_39}.

As wireless communications tuned increasingly to digital, the problem formulation broadened from emitter recognition to practical wireless security. This transition shifted attention from transient signatures to steady-state impairments and features observable in real communication frames~\cite{brik_wireless_2008_14}. Feature engineering became central, with researchers designing pipelines around amplitude, phase, frequency, oscillator imperfections, and other measurable artifacts, followed by classical \ac{ML} algorithms for classification and authentication~\cite{kennedy_radio_2008_29,liu_nonlinearity_2011_56,polak_wireless_2015_67}. This period was important not only because it improved performance in more realistic settings, but also because it established a recognizable \ac{RFF} workflow built around feature extraction, dimensionality reduction, and supervised decision-making.

In this context, a major turning point came with the adoption of \ac{DL}, especially convolutional architectures operating directly on raw \ac{IQ} samples or images generated from them~\cite{merchant_deep_2018_85}. Representation learning reduced reliance on handcrafted features, increasing performance in controlled experiments, accelerating research across a growing set of technologies and applications~\cite{peng_deep_2019_99}. At the same time, this progress exposed a deeper challenge: strong classification accuracy did not necessarily imply that models had learned stable transmitter-specific characteristics. Channel effects, receiver artifacts, session dependence, and dataset biases often contributed substantially to the performance, raising questions about what exactly was being fingerprinted and under which conditions the resulting systems remained dependable and trustworthy~\cite{al-shawabka_exposing_2020_123,hanna_wisig_2022}.

These observations have driven the field toward a more rigorous and application-oriented research paradigm, in which the focus extends beyond classification performance to the broader issues of reliability, trustworthiness, and ultimately dependability. Consequently, recent research increasingly emphasizes the use of shared datasets, reproducible evaluation, open-set recognition, adversarial robustness, and deployment realism~\cite{jagannath_comprehensive_2022}. This shift reflects an important recognition: for \ac{RFF} to serve as a dependable security primitive, it must operate under changing channels, unseen devices, heterogeneous receivers, and active attacks, not only in carefully controlled laboratory environments. In this sense, the central challenge facing \ac{RFF} today is no longer whether transmitter identification is possible, but how to make it reliable, interpretable, and operationally meaningful in the real world~\cite{al-hazbi_radio_2024_221}.

{\bf Contribution.} This paper presents a historical and technical review of \ac{RFF} from its origins in the early 1990s to its current state. Rather than treating the literature as a sequence of isolated methods, we organize the field around its major conceptual transitions, as illustrated in Fig.~\ref{fig:timeline}: from transient-based analysis to steady-state fingerprints, from handcrafted features to representation learning, and from laboratory experiments to real-world feasibility analyses. By tracing these shifts, we aim to clarify how the field developed, what assumptions shaped each phase, which limitations persisted across generations of methods, and where the most important open problems now lie. Overall, the objective of this paper is to provide a structured perspective on the evolution of \ac{RFF} and to help frame the next stage of research toward robust, reproducible, and practically deployable RFF-based identification systems.

{\bf Roadmap.} The remainder of this paper is organized as follows. Section~\ref{sec:paper_method} describes the paper selection methodology. Section~\ref{sec:radar_origins} traces the radar origins of \ac{RFF} and the transient-based paradigm. Section~\ref{sec:digital_communications} discusses the transition to digital communications and steady-state feature engineering. Section~\ref{sec:machine_learning} examines the \ac{ML} standardization period. Section~\ref{sec:deep_learning} analyzes the \ac{DL} transition and its implications for channel robustness and dataset realism. Section~\ref{sec:accuracy_to_credibility} addresses the shift from accuracy to real-world credibility and feasibility, covering open-set recognition, adversarial robustness, and real-world deployment. Finally, Section~\ref{sec:conclusion} concludes the paper.


\section{Papers Selection Methodology}
\label{sec:paper_method}
In this section, we present the list of papers considered throughout this work. To this end, we adopt a well-established systematic review methodology, namely Preferred Reporting Items for Systematic Reviews and Meta-Analyses (PRISMA)~\cite{prisma}, very frequently used by
researchers for rigorous and comprehensive literature reviews. In accordance with the PRISMA workflow, the selection process is organized
into four canonical phases: \emph{identification}, \emph{screening}, \emph{eligibility}, and \emph{inclusion}. In the \emph{identification}
phase, we queried the principal scientific repositories for \ac{RFF}, i.e., IEEE Xplore, ACM Digital Library, ScienceDirect, and Google Scholar, using a
set of domain-specific keywords, namely \emph{``Radio Frequency Fingerprinting''}, \emph{``RF fingerprinting''}, \emph{``Specific Emitter
Identification''}, \emph{``physical-layer device identification''}, and \emph{``transmitter identification''}, considering a time window spanning
from 1990 to 2025. This initial query returned a total of $755$ candidate papers. In the subsequent \emph{screening} phase, we removed duplicate
records and discarded contributions whose title and abstract revealed no substantive relevance to \ac{RFF}, e.g., works that merely mention the
topic in passing or address unrelated problems. During the \emph{eligibility} phase, we examined the full text of the remaining papers and excluded those that neither propose, evaluate, nor critically discuss a transmitter-identification methodology grounded in physical-layer hardware imperfections, along with non-peer-reviewed or non-archival entries lacking sufficient methodological detail. Finally, the works surviving this filtering process constitute the final set of $276$ papers analyzed for the chronicle discussed in this work.

\begin{figure*}[t]
 \centering
 \includegraphics[width=0.9\textwidth, angle = 0,trim = 0mm 60mm 0mm 0mm]{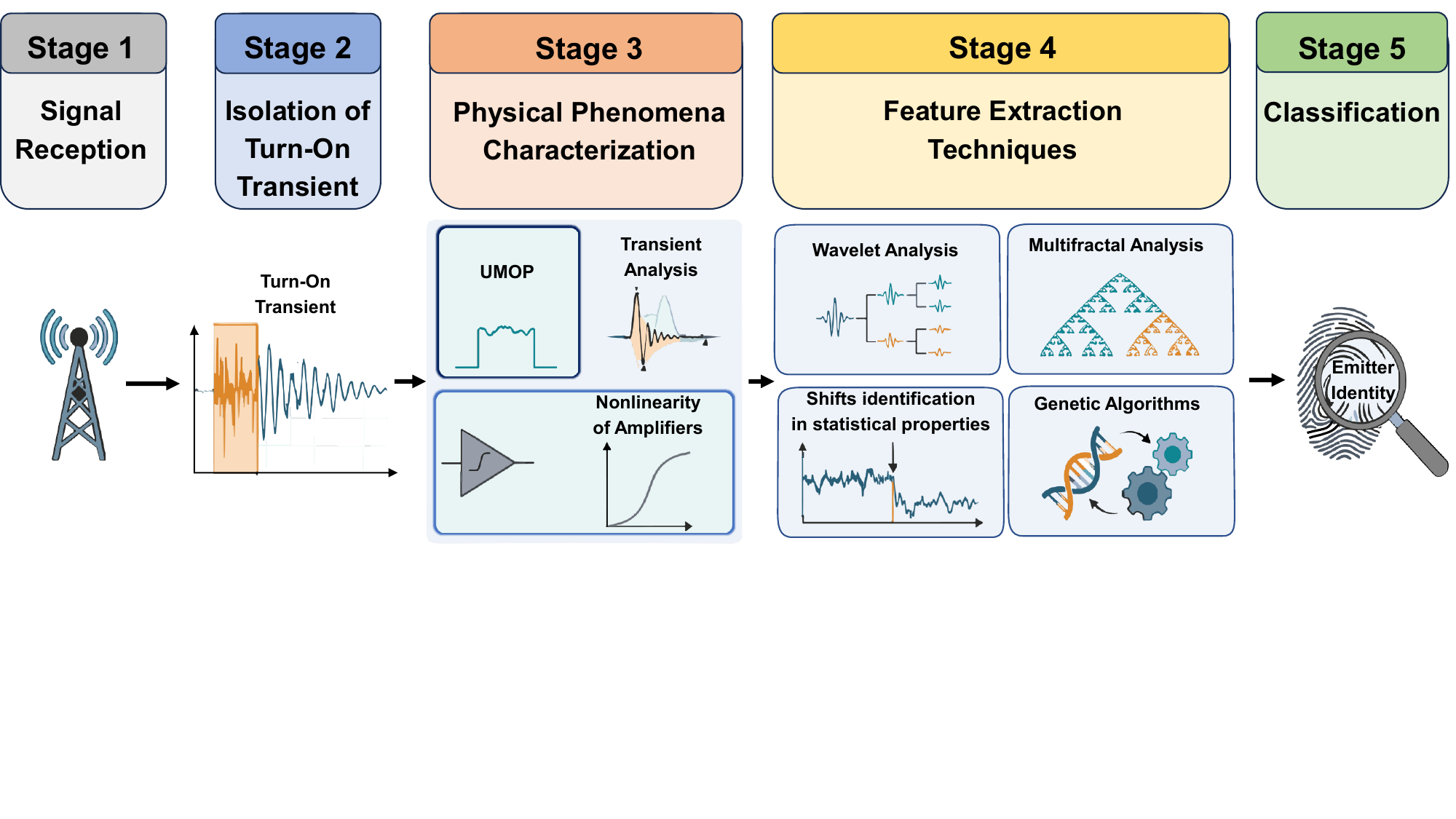}
 \caption{Early transient-based \ac{RFF} techniques.}
 \label{fig:transient}
\end{figure*}
 
\section{The Radar Origins and the ``Transient'' feature}
\label{sec:radar_origins}

This section traces the earliest roots of \ac{RFF} in the radar domain, where the concept of \ac{SEI} first emerged, and examines how transient-based signal features became the dominant fingerprinting paradigm during the field's foundational decade.

The earliest contribution on \ac{RFF} identified in the literature dates back to 1993, with the work by Lawrence E. Langley titled ``\ac{SEI} and Classical Parameter Fusion Technology''~\cite{langley_specific_1993_6}. The seminal paper introduces the problem of \ac{SEI} while acknowledging the difficulty of transmitter identification based solely on parameters such as frequency and pulse repetition interval. To address this limitation, the proposed solution leverages \ac{UMOP}, the unavoidable and unique features of a signal's pulse waveform that can be used to identify a specific emitter. While the reference scenario considers identifying radar sources (pulse-based signaling), the author claims that the solution is general and applicable to other types of signals.

{\bf The signal transient feature.} The period between 1993 and 2001 experienced more papers characterized by the idea of extracting the features from the signal {\em transients} to identify the transmitting source. The transient can be defined as the relatively short phase through which a transmitter goes when activated and when the \ac{RF} power or the carrier frequency changes. Different techniques have been considered to uniquely identify the transmitter exploiting the transients, spanning between wavelet analysis~\cite{toonstra_radio_1996_2}, genetic algorithms~\cite{toonstra_transient_1995_7}, multifractal analysis~\cite{shaw_multifractal_1997_5}, and Bayesian change point detection~\cite{ureten_bayesian_1999_4}. The physical phenomena considered for the classification of the emitter are mainly three: (i) \ac{UMOP}, (ii) transient analysis, and finally, (iii) nonlinearity of amplifiers, as depicted in Fig.~\ref{fig:transient}. Inspired by the early work in 1993~\cite{langley_specific_1993_6}, several papers exploited \ac{UMOP} to achieve emitter identification~\cite{li_combining_2009_20}. In parallel, the nonlinear characteristics of power amplifiers were investigated as an independent fingerprint source~\cite{yu_feature_2009_22}, where the root idea was that each amplifier exhibits unique nonlinear characteristics due to variations in design and operating conditions, thereby generating unique identifiers for \ac{RFF}. In this context, it is worth mentioning that these first 10 years were characterized by the belief that the ``turn-on'' transient constituted the only reliable fingerprint. This was partially correct but, at the same time, limiting: transients are inherently short, hard to capture, require high sampling rates, and they are not available during the vast majority of the communication. Consequently, all the proposed techniques for processing, detecting, and uniquely identifying transmitters based on this feature were attempts to extract information from a difficult-to-observe phenomenon.

\section{RFF for Digital Communications}
\label{sec:digital_communications}
This section examines the transitioning of \ac{RFF} from the military radar domain into civilian wireless networks, tracing the shift from transient analysis to steady-state signal features and the emergence of feature engineering as the central methodological effort.

\begin{figure*}
 \centering
 \includegraphics[width=0.95\textwidth, angle = 0,trim = 0mm 50mm 0mm 0mm]{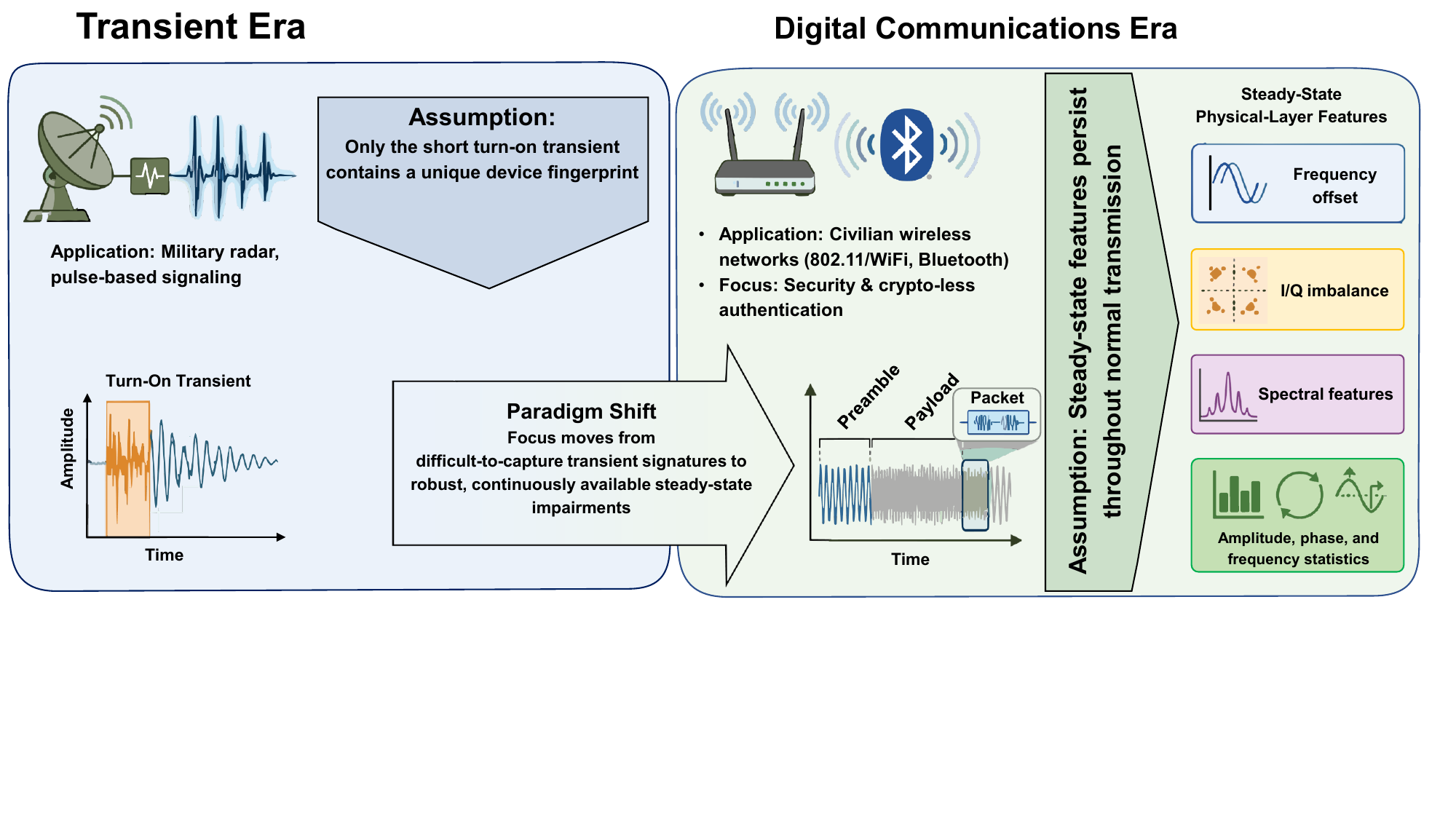}
 \caption{Shifting from transient to steady-state feature extraction.}
 \label{fig:trans_to_digital}
\end{figure*}

The transient-based paradigm, while well-suited to the military radar context in which \ac{RFF} originated, became increasingly inadequate as the wireless landscape evolved. As summarized in Fig.~\ref{fig:trans_to_digital}, by the mid-2000s, IEEE 802.11 networks had proliferated into homes and offices, Bluetooth had established itself as the dominant short-range connectivity standard (e.g., connecting headsets and keyboards), and the prospect of crypto-less and \ac{PHY}-layer device authentication had gained practical relevance beyond the military domain. Thus, the transient paradigm was about to be challenged, not by a theoretical argument, but by a change of application domain.

{\bf Wireless Security.} In this context, the turning point can be traced back to~\cite{ureten_wireless_2007_39}, i.e., the first work applying \ac{RFF} to digital communications. The paper, titled ``Wireless Security through RF Fingerprinting'' represents a landmark not because it introduced a new technique but because it changed the application scenario. Indeed, the paper still relies on transient analysis as its core method, while highlighting that amplitude characteristics during startup offer richer discriminative content than phase characteristics. Nevertheless, while prior work focused on extracting turn-on signatures from radios, this was the first time the authors looked at the problem of identifying wireless network devices based on their inherent hardware imperfections. This contribution effectively repositions \ac{RFF} as a \ac{PHY}-layer security primitive for civilian wireless networks, a direction subsequently reinforced by~\cite{candore_robust_2009_31}.

{\bf Steady-state Features}. The definitive shift from transient analysis to steady-state features happened in 2008. The contribution in~\cite{brik_wireless_2008_14} demonstrated that \ac{PHY}-layer features extracted from IEEE 802.11 frames---including frequency offset, \ac{IQ} imbalance, and spectral features---could be used to identify individual network cards with high accuracy. This paper was the first to consider commercial off-the-shelf hardware, a realistic network security scenario, and the application of \ac{SVM} for \ac{RFF}. 

{\bf Features Engineering.} Subsequently, the period between 2010 and 2015 was characterized by a growing effort on feature extraction and to move \ac{RFF} beyond proof-of-concept. Herein, the approach RF-DNA (Radio Frequency Distinct Native Attribute) framework~\cite{paillier_rf-dna_2007_23}, developed across multiple papers, provided a methodology for extracting second-order statistical descriptors, i.e., mean, standard deviation, skewness, and kurtosis, from amplitude, phase, and frequency of radio signals. Applied first to IEEE 802.11 WiFi devices~\cite{cobb_physical_2010_28}, and subsequently to ZigBee nodes\cite{patel_improving_2014_57}, RF-DNA represents a preliminary attempt to systematize the feature engineering pipeline in \ac{RFF}. Finally, the contribution in \cite{patel_comparison_2014_47} explored RF-DNA performance across high-end and low-end receivers, an early acknowledgment that the hardware itself could influence fingerprinting accuracy---a problem that continues to affect the topic still today.

\section{Standardization of the RFF Workflow: the Machine Learning period}
\label{sec:machine_learning}
This section discusses how the \ac{RFF} community converged on a standardized \ac{ML}-driven pipeline, covering device authentication, feature engineering, measurement realism, and the security-motivated applications that defined the field between 2015 and 2018.

\begin{figure*}[t]
 \centering
  \includegraphics[width=0.9\linewidth]{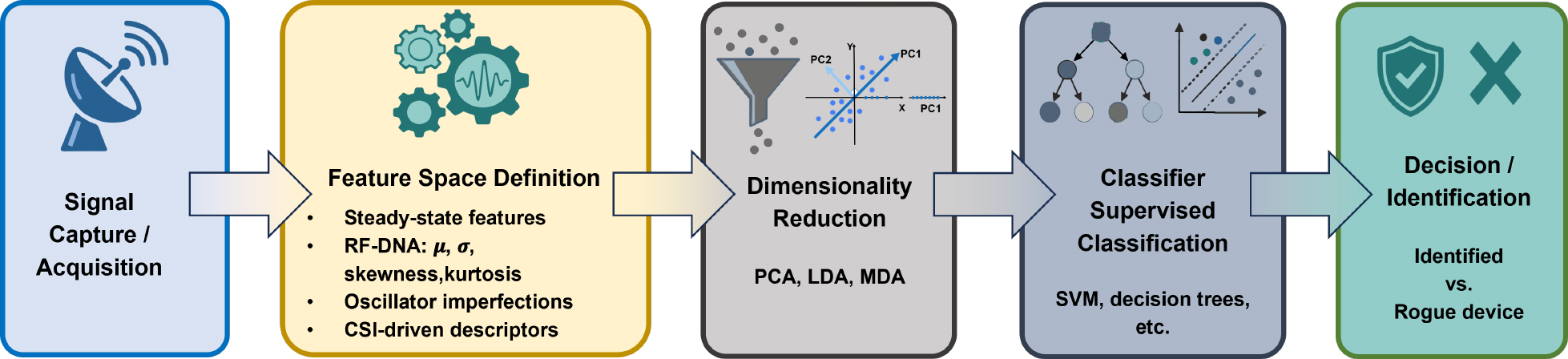}
 \caption{Standardized ML-based security-oriented and classifier-driven pipeline.}
 \label{fig:machine_learning}
\end{figure*}

By 2015, \ac{RFF} had largely converged on adopting information from the ``steady-state'': although transients remained attractive, the community increasingly focused on what was available in preambles, payload segments, and stable RF impairments that could be extracted without ultra-high sampling. This period was characterized by convergence toward a repeatable \ac{ML} pipeline despite the search for signal features, including defining the feature space, regularization, and classifier selection. A key signal of this consolidation is the line of work subsequent to RF-DNA~\cite{paillier_rf-dna_2007_23} shown in Fig.~\ref{fig:machine_learning}, which has moved toward an explicitly security-oriented, classifier-driven framework.

{\bf Device Identification.} Here, the authors in \cite{reising_authorized_2015_49} formalized the concept of ``identified vs. rogue'' discrimination using dimensionally reduced RF-DNA fingerprints, positioning \ac{RFF} as a practical identification primitive rather than a lab curiosity~\cite{reising_authorized_2015_49}. In parallel, the pool of features broadened toward impairments that were both measurable and stable under standard receiver chains. The work on oscillator imperfections~\cite{polak_wireless_2015_67} is an interesting case: it treats \ac{RF} impairments as a structured source of identity information that can be used for classification under realistic conditions. At the same time, another important aspect was emerging: if \ac{RFF} is to be used for security, adversaries will respond. This aspect becomes explicit in the study of users who actively fake fingerprints via artificial distortion, stressing that ``high accuracy'' in benign conditions is not the same as robustness in a hostile environment~\cite{polak_identification_2015_53}.

{\bf Feature engineering.} The rise of \ac{ML} in \ac{RFF} also coincided with \ac{RFF}’s expansion beyond the canonical WiFi security topic into adjacent application domains, some of which were not originally framed as \ac{RFF}. For instance, \ac{CSI}-driven work in 2015 demonstrated how channel-related measurements can be leveraged for fingerprinting-like inference and tracking, even though it complicates what exactly is being ``fingerprinted'' (device, channel, or both)~\cite{chapre_csi_2015_42}. Similarly, dual-link designs for wearables showed the community experimenting with system-level observation strategies to stabilize and enrich fingerprints in realistic deployments~\cite{revadigar_dlink_2015_50}. The period also saw continued diversification into specialized settings, from infrastructure contexts to additional \ac{RF} modalities, often with the same recipe: engineering a feature vector, reducing it, and finally classifying.

{\bf Measurement Realism and Sensitivity.} In the following years (2016–2017), two themes emerged: (i) real measurement scenarios and (ii) sensitivity analysis. The work in~\cite{vo-huu_fingerprinting_2016_74} highlighted how \ac{SDR} enabled more controlled experimentation while also exposing how implementation details and capture pipelines can shape the resulting ``fingerprint''. As for sensitivity, work such as the assessment of \ac{CFO} impact on RF-DNA classification made explicit what many earlier papers only implied: performance can hinge on seemingly secondary synchronization and front-end effects, and the measurement chain must be treated as part of the system and not simply as a neutral window into transmitter identity~\cite{wheeler_assessment_2017_72}. 

{\bf Security-motivated applications.} Security-motivated applications widened further as well. The case of FBSleuth~\cite{zhuang_fbsleuth_2018_84} is illustrative: it frames \ac{RFF} as a forensic tool to detect fake cellular base stations, which implicitly raises the bar from ``classify known devices'' to ``withstand manipulation and deception in the field''. Complementary protocol-layer constructions also appeared, pairing \ac{RFF} with authentication mechanisms, e.g., lightweight one-time password schemes, that treat the fingerprint as one component in a broader security design~\cite{chen_lightweight_2018_87}. By this point, the field had accumulated enough ``working demos'' that the natural next question was not whether \ac{RFF} can work, but under what assumptions it keeps working and which assumptions are the strongest and hardest to keep in the wild.

\begin{figure}
 \centering
 \includegraphics[width=\columnwidth]{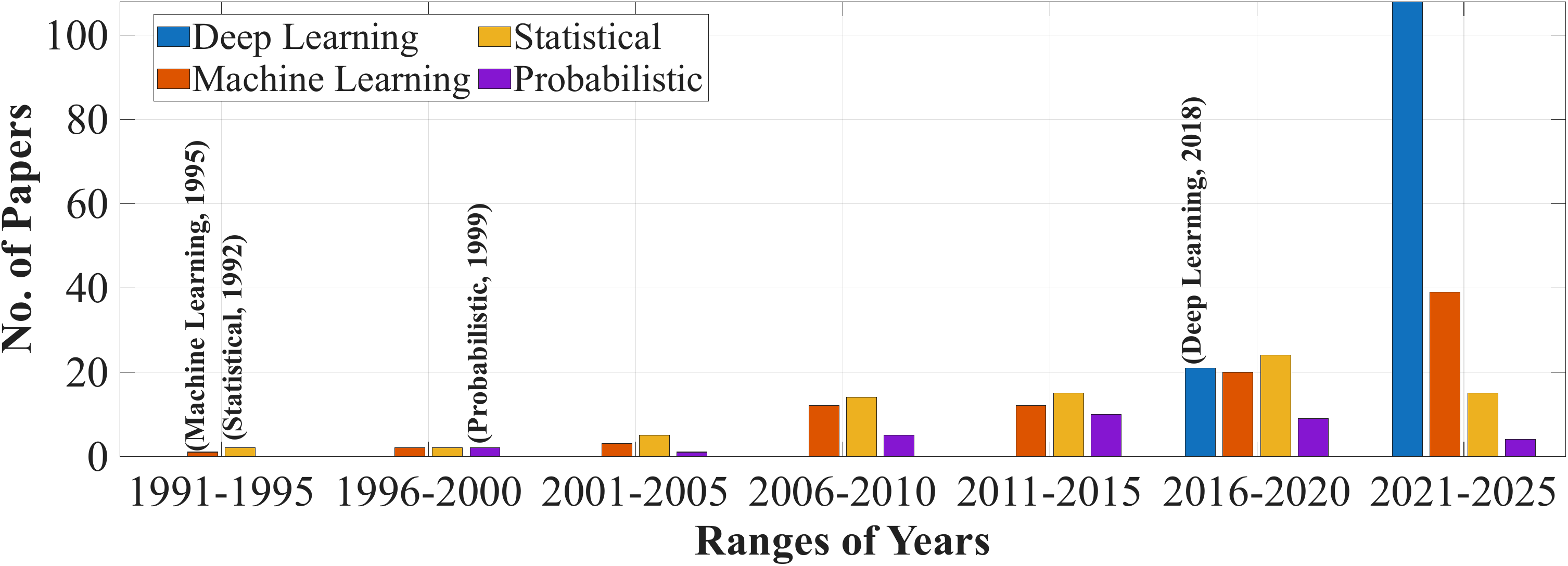}
 \caption{Evolution of classification techniques over the years.}
 \label{fig:pivot_year}
\end{figure}

{\bf The 2018 pivot.} As illustrated in Fig.~\ref{fig:pivot_year}, the hinge year is 2018. On the one hand, it marked the peak of confidence in the classical pipeline: hybrid extraction/classification designs continued to refine the engineered-feature paradigm and its decision logic~\cite{peng_design_2018_86}, and \ac{ML}-based transmitter identification frameworks were presented as practical tools~\cite{youssef_machine_2018_90}.
On the other hand, the same year witnessed the emergence of \ac{DL} for \ac{RF} device fingerprinting, with multiple contributions arguing that end-to-end representation learning from raw signal data could outperform handcrafted feature pipelines while substantially reducing the need for domain-specific engineering~\cite{jafari_iot_2018_88}. Moreover, as Fig.~\ref{fig:pivot_year} shows, \ac{DL} adoption grew rapidly thereafter, overtaking all other paradigms by 2021--2025, when it accounts for the large majority of published work, while statistical and \ac{ML} methods have persisted but stabilized.

{\bf Limits of the classical pipeline.} Specifically, classical \ac{ML} had become disciplined and effective, still relying on crafted features, small private datasets, and evaluation setups that rarely separated transmitter identity from receiver/channel idiosyncrasies~\cite{vo-huu_fingerprinting_2016_74}.
The legacy of this period, therefore, was not a single classifier or feature family, but it was the standardization of an evaluation methodology and the realization of its limits. The field had learned how to build pipelines that look strong in controlled settings, and it had begun to articulate threats (active distortion, protocol manipulation, hardware/estimation sensitivity) that could break those pipelines in practice~\cite{polak_identification_2015_53}. This combination established the premises for the \ac{DL} transition, in which the central expectation shifts from accurate classification to credible and verifiable representation of transmitter identity.

\section{From Feature Engineering to Representation Learning}
\label{sec:deep_learning}
This section analyzes the rapid transition to \ac{DL} that began in 2018, examining both the performance gains enabled by end-to-end representation learning and the structural weaknesses it introduced, including channel memorization, architectural fragmentation, and adversarial susceptibility.

\begin{figure*}
 \centering
 \includegraphics[width=0.9\textwidth, angle = 0,trim = 0mm 50mm 0mm 20mm]{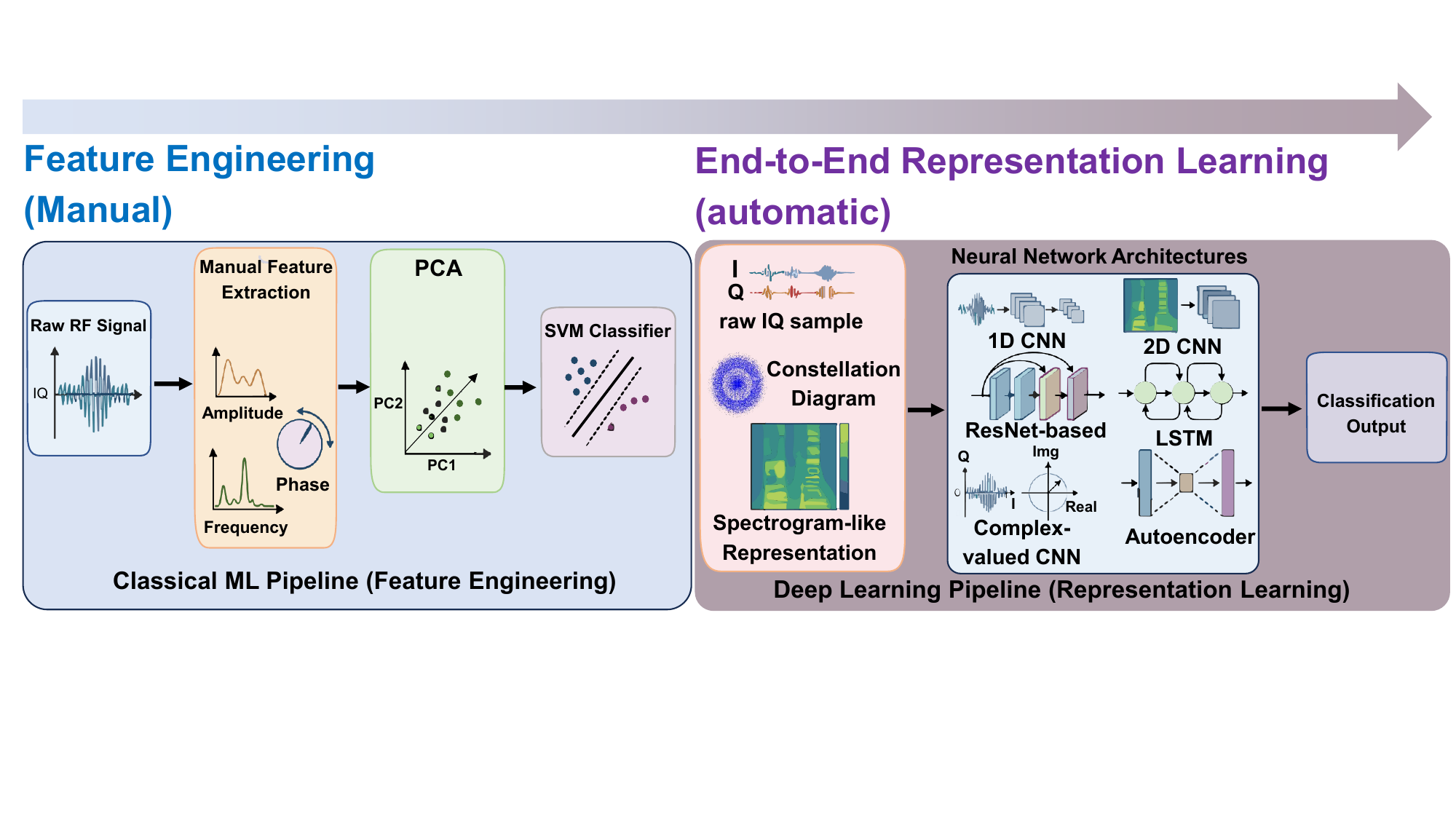}
 \caption{\ac{RFF}: from (manual) feature engineering to end-to-end representation learning.}
 \label{fig:representation_learning}
\end{figure*}

As discussed in the previous section and shown in Fig.~\ref{fig:representation_learning}, 2018 marked an abrupt methodological discontinuity in \ac{RFF} research. In a single year, \acp{CNN} became the de facto standard of the field. 
Specifically, \ac{CNN} adoption, virtually absent before that year, expanded to tens of publications by the 2021--2025 period, establishing \ac{DL} as the dominant paradigm within a remarkably short timeframe. This section examines the technical and contextual factors that drove this transition, the performance gains it enabled, and the structural vulnerabilities it introduced.

{\bf Deep Learning Transition.} The root cause behind the wide adoption of \ac{CNN} for \ac{RFF} was not a theoretical insight about \ac{RF} hardware. We can consider it as an architectural accident: the raw IQ samples, the most natural representation of a received radio waveform, turn out to have exactly the spatial and temporal structure that convolutional filters are designed to exploit.
Rather than engineering feature vectors through amplitude, phase, and frequency extraction followed by \ac{PCA} and \ac{SVM} classification, a 1D-\ac{CNN} trained directly on \ac{IQ} samples could learn discriminative representations with substantially less manual intervention.
Consequently, Merchant et al.~\cite{merchant_deep_2018_85} demonstrated this for cognitive radio networks in 2018; Jafari et al.~\cite{jafari_iot_2018_88} showed it for \ac{IoT} devices in the same year; and Ding et al. ~\cite{ding_specific_2018_95} applied convolutional identification to specific emitter identification, confirming that the approach generalized across application contexts. Moreover, the authors in~\cite{riyaz_radio_2018_91} examined \ac{CNN}-based fingerprinting for IEEE 802.11 devices and became an early reference point for the paradigm. The message was consistent across all four: end-to-end learning from raw signals outperforms handcrafted feature pipelines, at least in the controlled conditions in which these papers were evaluated. Moreover, \ac{DL} also promised to resolve a bottleneck that had constrained the classical pipeline for years: the dependence of classification performance on feature engineering choices that required expert knowledge and were difficult to verify.

{\bf Channel Robustness.} Almost immediately, \ac{DL} faced a complication. The high classification accuracy reported by early \ac{DL}-based \ac{RFF} systems — often above 99\% — was consistently achieved under near-ideal experimental conditions: training and test data collected in the same environment, on the same day, using the same receiver hardware. Here, what the models were learning was not entirely the transmitter's fingerprint, but a significant portion of the ``signal'' being classified was the fingerprint of the channel, i.e., the specific propagation environment between transmitter and receiver at the moment of data collection.
Although not a new concern - receiver front-end effects had been identified as a confounding factor as early as 2012~\cite{rehman_analysis_2012_45}, and prior work on oscillator imperfections had deliberately framed device identity in terms of stable, estimable parameters~\cite{polak_wireless_2015_67} - the magnitude of the problem in the \ac{DL} context proved qualitatively different. Specifically, a \ac{CNN} trained on raw IQ samples can not distinguish between a consistent transmitter impairment and a consistent channel effect since both appear as structured deviations from the ideal waveform, and both will be exploited if they improve training loss. This fundamental ambiguity was systematically exposed by Al-Shawabka et al.~\cite{al-shawabka_exposing_2020_123}, which systematically dissected the impact of the wireless channel on radio fingerprinting accuracy and demonstrated that performance degraded sharply when channel conditions changed between training and testing. The implication is clear: much of the accuracy credited to transmitter fingerprinting was actually channel memorization. Furthermore, Restuccia et al.~\cite{restuccia_deepradioid_2019_97} attempted to address this directly in 2019, proposing channel-resilient optimization for \ac{DL}-based fingerprinting, which included 500 ADS-B and 500 WiFi devices (DARPA RFMLS dataset). In parallel, Sankhe et al.~\cite{sankhe_no_2019_102} introduced hardware impairment fingerprinting through \ac{DL} of physical-layer features, explicitly attempting to separate device identity from channel effects. These papers represent the first systematic acknowledgment that the channel problem was a central design challenge.

{\bf Datasets Scale.} The DARPA RFMLS program is the institutional pivot point of this period for large-scale \ac{RFF} evaluation. The release of public datasets in 2019, e.g., the DeepRadioID dataset~\cite{restuccia_deepradioid_2019_97}, the large-scale WiFi and ADS-B collections~\cite{gritsenko_finding_2019_100}, and the 400 GB corpus described in~\cite{al-shawabka_exposing_2020_123}, was the first time that results from different research groups could be compared on common ground. The authors in~\cite{jian_deep_2020_120} conducted a comprehensive experimental study on these datasets by running multiple \ac{DL} architectures on a unified corpus of over 10,000 devices, revealing that the "massive experimental study" label was well-earned and that performance varied considerably across architectures, preprocessing, and evaluation protocols. The result was simultaneously encouraging, i.e., \ac{DL} scaled to large device populations, and concerning, i.e., the accuracy advantages reported in lab-scale studies did not always survive at scale or across sessions.

{\bf Architectural Diversification.} Between 2019 and 2022, research on \ac{CNN} architectures for \ac{RFF} expanded rapidly. One-dimensional \ac{CNN} for temporal IQ processing, two-dimensional \ac{CNN} for spectrogram analysis, complex-valued \ac{CNN} for exploiting the full IQ structure~\cite{wang_efficient_2021_138}, ResNet-based architectures, \ac{LSTM} networks for sequential dependency modeling~\cite{tyler_analysis_2022_151}, and autoencoder-based approaches for unsupervised representation learning~\cite{huang_deep_2022_149} all appeared within a few years of each other. The trend reflected both the significance of the approach and a growing recognition that no single architecture was universally optimal. Simultaneously, the application space continued to expand. \ac{LoRa} fingerprinting via \ac{DL} emerged as a well-developed sub-field, with ~\cite{shen_radio_2021_139} demonstrating \ac{CNN}-based identification on \ac{LoRa} signals and subsequently addressing scalability and channel robustness ~\cite{shen_towards_2022_162}. ADS-B fingerprinting, an application where the absence of encryption makes \ac{RFF} a practical security primitive, attracted growing attention ~\cite{restuccia_deepradioid_2019_97}. Satellite authentication via \ac{DL} was first demonstrated by PAST-AI~\cite{oligeri_past-ai_2023_180}, which performed \ac{PHY}-layer authentication of Iridium transmitters. Each new application domain brought its own channel model, device population, and protocol structure, testing whether the \ac{DL} approach was genuinely general or merely well-adapted to the wireless lab environment where it had been developed.

{\bf The Adversarial Perspective.} As \ac{DL} raised accuracy, it also raised the possibility of adversarial manipulation. The adversarial literature of this period is characterized by two concerns that remain distinct but related. The first is the obfuscation attack: a device that actively manipulates its transmissions to masquerade as another. The authors in~\cite{polak_identification_2015_53} had identified this threat conceptually in 2015, but \ac{DL} classifiers are potentially more vulnerable to carefully crafted perturbations than \ac{SVM} trained on hand-crafted features, because they optimize over a high-dimensional input space with many exploitable directions. The authors in~\cite{sun_robustness_2022_161} examined the adversarial robustness of \ac{DL}-based \ac{SEI} under adversarial attacks in 2022, and the field subsequently initiated a systematic evaluation of this threat. The second concern is the mimicking adversary: an attacker who replays or synthesizes signals that fool the classifier without needing to understand the model. The authors in~\cite{reising_rf_2023_188} examined fingerprint-based identity verification in the presence of SEI-mimicking adversaries, showing that designing robust discriminators under active attack was substantially harder than classification in benign conditions.

{\bf Open Challenges.} By the end of this period, a structural pattern had re-emerged at a larger scale: \ac{DL} had resolved the feature engineering bottleneck inherited from the classical \ac{ML} era, but had not resolved the evaluation problem. Laboratory accuracy remained impressively high, while generalization across sessions, receivers, and channels remained inconsistent. As for the dataset infrastructure, while vastly improved by the DARPA program, it was still concentrated in WiFi and ADS-B, with limited representation of multipath environments and essentially no standardized adversarial benchmarks. The critical observation in ~\cite{fadul_rf-dna_2019_109} that ``despite over two decades of \ac{RFF} research, all but four papers have assumed an AWGN channel model'' is still consistent even if \ac{DL} provided more powerful classifiers.

\begin{figure}
 \centering
 \includegraphics[width=\columnwidth]{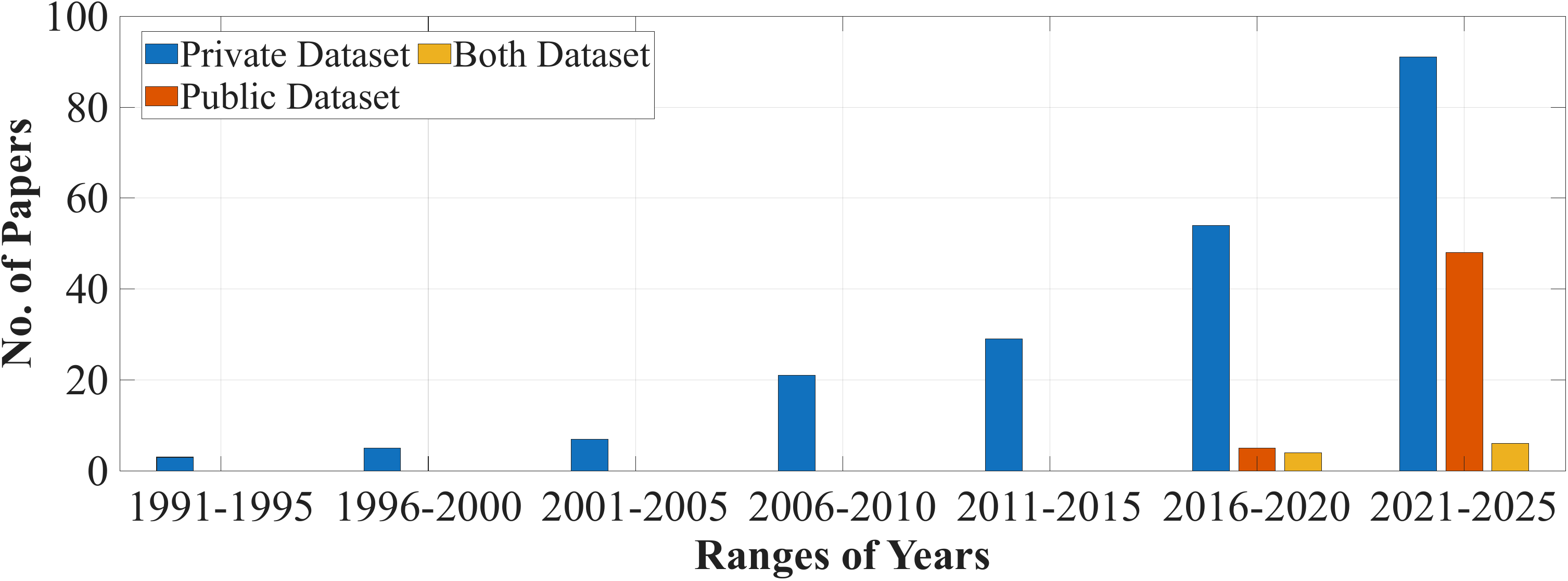}
 \caption{Evolution of dataset types over the years.}
 \label{fig:dataset_over_time}
\end{figure}

\section{From Accuracy to Credibility}
\label{sec:accuracy_to_credibility}
This section describes the latest and current phase of \ac{RFF} research, in which the community has shifted focus from maximizing classification accuracy to establishing credibility under realistic conditions, addressing open-set recognition, adversarial robustness, dataset realism, and real-world deployment across a growing range of wireless technologies.

\begin{figure*}
 \centering
 \includegraphics[width=\textwidth, angle = 0,trim = 10mm 40mm 10mm 10mm]{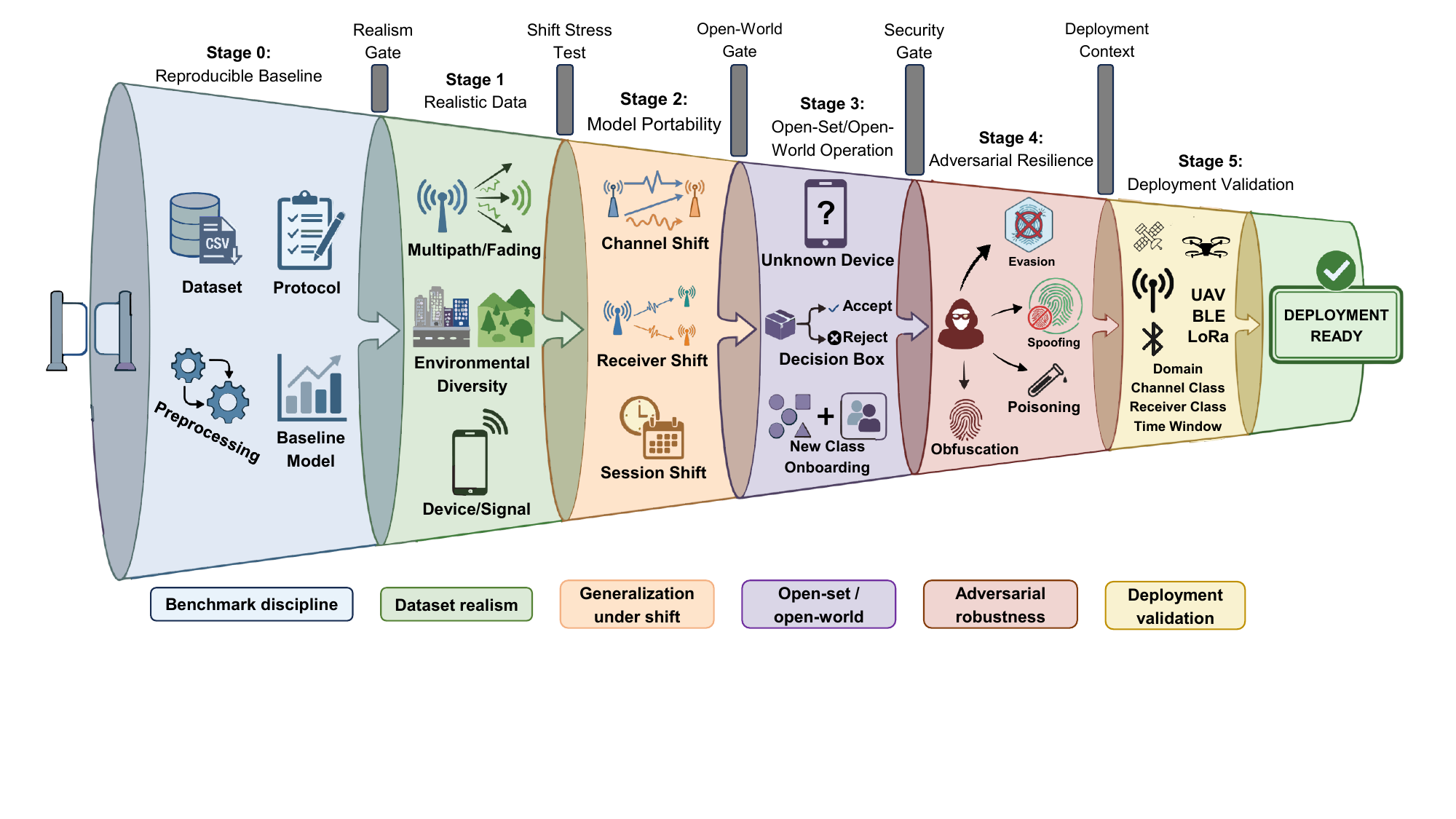}
 \caption{From laboratory to deployment: credibility tunnel for \ac{DL} model evaluation.}
 \label{fig:credibility}
\end{figure*}

By the early 2020s, \ac{RFF} had already demonstrated that \ac{DL} could deliver impressive classification performance. The central question was no longer whether \acp{NN} could discriminate devices under controlled conditions, but whether the system could produce results that remained meaningful outside the laboratory. Thus, the current phase of \ac{RFF} is defined less by the design of a single new classifier than by a shift in focus: from accuracy maximization on convenient datasets to credibility under deployment-relevant constraints. Herein, Fig.~\ref{fig:credibility} shows the evaluation progression to include public benchmarks, open-set recognition, cross-domain robustness, adversarial resilience, and long-term deployment effects, which are becoming the core of the current state of the art.

{\bf Dataset Realism.} One visible sign of this transition is the growing importance of shared datasets and reproducible evaluation. The field had already crossed a threshold in 2019 with the first large-scale public dataset~\cite{restuccia_deepradioid_2019_97}, but from 2022 onward, its influence became much more pronounced. Scientific studies increasingly relied on datasets such as RFMLS-NEU~\cite{muller_sensitivity_2024_224}, WiSig~\cite{hanna_wisig_2022}, ORACLE~\cite{sankhe_oracle_2019}, and technology-specific collections for \ac{LoRa}, \acp{UAV}, ZigBee, satellites, and ADS-B. This did not eliminate the adoption of private data (and small ad-hoc datasets), but it changed the field’s standards. Once multiple groups could test on the same large dataset, claims became easier to compare, and weaknesses that had been hidden by isolated, custom-built testbeds became harder to ignore. This new trend is captured especially well by recent work explicitly arguing that \ac{RFF} now needs better data, not bigger models~\cite{bothereau_why_2025_275}. This evolution is further illustrated in Fig.~\ref{fig:dataset_over_time}, which tracks the distribution of papers by dataset type across five-year intervals from 1991 to 2025. Specifically, private datasets dominated for the first three decades, with public-dataset studies remaining negligible before 2016. A clear inflection point emerges between 2021 and 2025, when public dataset adoption rose sharply, signaling that reproducibility and cross-study comparability have become explicit priorities for the community.

{\bf Open-Set Recognition.} The shift toward deployment-oriented evaluation has been accompanied by a transition to more challenging learning formulations. Earlier work had mostly assumed a closed world: every transmitter seen at test time belonged to a class already represented during training. In practice, this assumption is too strong for security applications, where the system must often decide whether a device is known, unknown, suspicious, or only partially represented in the training set. As a result, open-set and open-world scenarios have gained substantial traction. Open-set radar \ac{SEI} appears in~\cite{jing_dynamic_2025_264}, open-set authentication in noisy channels is addressed in~\cite{huang_radio_2024}, and a broader open-world formulation is developed in~\cite{han_open-world_2025_235}. Closely related are few-shot and class-incremental approaches, which recognize that practical systems must absorb new emitters over time without full retraining. Representative examples include deep metric ensemble learning~\cite{wang_few-shot_2022_157}, limited-sample and meta-learning approaches~\cite{yang_specific_2022_163}, masked-autoencoding for learning from scarce labels~\cite{huang_deep_2022_149}, class-incremental \ac{SEI}~\cite{li_fscil-sei_2025_251}, and few-shot channel-robust identification~\cite{fu_channel-robust_2025_241}. Collectively, these works reframe \ac{RFF} as an evolving recognition problem rather than a static closed-set classifier.

{\bf Adversarial Robustness.} As \ac{RFF} matured into a candidate transmitter identification primitive, adversarial robustness emerged as a central evaluation criterion.
It became necessary to investigate how an attacker might evade, poison, or manipulate the classifier. In this context, recent work focuses on adversarial examples against \ac{DL}-based RFF~\cite{papangelo_adversarial_2024_199,sun_robustness_2022_161}, white-box anti-jamming and robustness schemes~\cite{hou_prototype-guided_2025_261}, poisoning-attack mitigation~\cite{baldini_mitigation_2025_259}, and practical fingerprint obfuscation on Bluetooth Low Energy (BLE) devices~\cite{givehchian_practical_2024_219}. Moreover, related work examines the mimicking of adversaries~\cite{reising_rf_2023_188} and adversarial attacks on domain-specific systems, such as \ac{LoRa}~\cite{sagduyu_adversarial_2023}. This line of research is especially important because it exposes a recurring tension in the field: the same high-dimensional representations that make \ac{DL} models powerful also create new attack surfaces. Thus, the present phase of \ac{RFF} is characterized by an important inversion with respect to prior phases. Earlier periods treated robustness as something that could be added after accuracy was achieved, while current work increasingly recognizes that accuracy without resilience does not achieve actual security.

\begin{figure}
 \centering
 \includegraphics[width=\columnwidth]{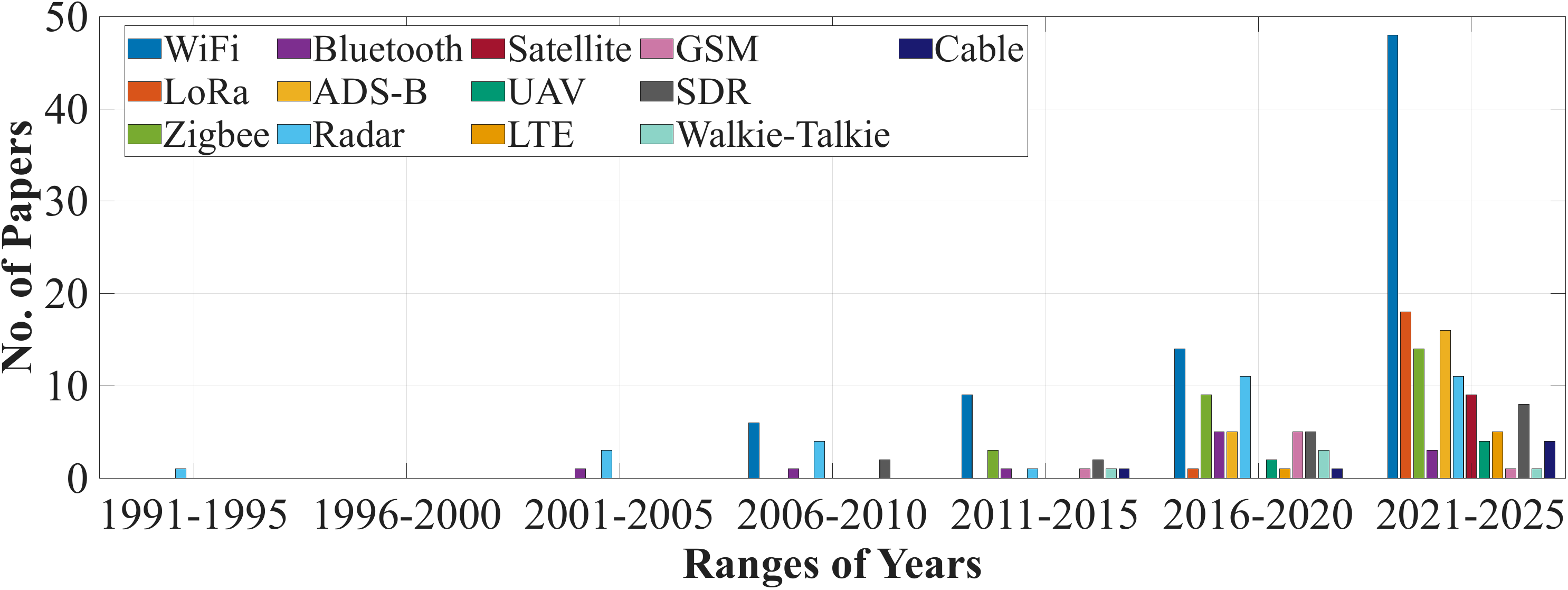}
 \caption{Evolution of technological scenarios over the years.}
 \label{fig:tech_over_time}
\end{figure}

{\bf Toward Real-World Deployment.} At the same time, the number of application scenarios increased significantly, as captured in Fig.~\ref{fig:tech_over_time}, indicating that \ac{RFF} is no longer centered solely on WiFi.
While WiFi has remained the single most studied technology throughout the entire period, the 2021--2025 interval saw a marked diversification, with \ac{LoRa}, ADS-B, \ac{UAV}, ZigBee, and satellite-based systems all reaching comparable publication volumes. \ac{LoRa} has become one of the most active subdomains~\cite{zhang_rapid_2026}, while ADS-B remains a major application because of its practical security relevance and the availability of large datasets~\cite{zhang_real-world_2024}. Satellite authentication has emerged from a niche topic into a more concrete research direction, spanning PAST-AI~\cite{oligeri_past-ai_2023_180}, FadePrint~\cite{oligeri_fadeprint_2024_210}, broader satellite network evaluations~\cite{cui_performance_2023_181}, and resilient satellite transmitter fingerprinting~\cite{smailes_watch_2023}. \ac{UAV} identification has similarly developed into a recognizable cluster~\cite{tian_optimized_2024}, while ZigBee, Bluetooth, LTE/5G, Radio-Frequency Identification (RFID), Automatic Identification System (AIS), and even power-line continue to extend the field's boundaries~\cite{irfan_device_2025}, with the latter demonstrating that hardware-induced imperfections in \ac{PLC} transmitters are sufficiently distinctive for \ac{CNN}-based device fingerprinting over wired infrastructure. The diversity of these deployment scenarios suggests that \ac{RFF} is adaptable, but each technology brings its own channel conditions, signal structure, adversarial model, and notion of what counts as a {\em stable fingerprint}.
Compounding this challenge, recent work has shown that fingerprint stability itself cannot be taken for granted: hardware-level phenomena such as temperature variation~\cite{temp_rff_gu2024}, device reboots~\cite{dat_Alhazbi_2023}, and \ac{FPGA} reconfigurations can alter the learned representation in ways that undermine identification reliability~\cite{irfan_cps-sec_2025}. 
Specifically, \cite{irfan_cps-sec_2025} demonstrated that \ac{FPGA} image reloads on a USRP X310 induce probabilistic fingerprint mutations at every new communication session. This reveals that a transmitter is better described by a set of distinct fingerprint \emph{states} rather than a single persistent one, and that reliable authentication requires observing at least 60\% of its ground-truth measurements. Building on this, \cite{static_sig_oligeri_2026} extended the analysis to both transmitter and receiver sides, showing that such mutations reduce to a two-state probabilistic process driven by residual phase and timing synchronization errors introduced by \ac{SDR} clocks. While external clock referencing eliminates the effect entirely, such synchronization is impractical in real-world wireless deployments, leaving fingerprint variability an open challenge.

Moreover, a related source of instability concerns the receiver itself, since its front-end introduces impairments that become entangled with the transmitter signature in the learned representation. This was systematically characterized in~\cite{rehman_analysis_2012_45}, which demonstrated that receiver front-end effects constitute a confounding factor in fingerprint extraction. Subsequently, the work in~\cite{patel_comparison_2014_47} compared high-end and low-end receivers and showed measurable accuracy degradation with lower-quality hardware. While an assessment of \ac{CFO} impact on \ac{RF}-DNA classification~\cite{wheeler_assessment_2017_72} made explicit that the measurement chain must be treated as part of the system rather than a neutral window onto transmitter identity. More recently, the WiSig dataset~\cite{hanna_wisig_2022} was explicitly designed to benchmark receiver- and channel-agnostic fingerprinting at scale, confirming that cross-receiver generalization remains one of the most challenging open problems in the field.

In addition, the adversarial dimension of \ac{RFF} has begun to receive explicit attention, with studies examining how fingerprinting systems can be evaded or manipulated rather than merely circumvented by passive channel variation~\cite{Tianya_exp_attack_2025, Zhisheng_spoofing_via_rff}.
In this context, the current period marks the beginning of a more demanding phase in which \ac{RFF} is asked to justify itself under conditions that matter most: unknown devices, changing channels, heterogeneous receivers, active attackers, and realistic operational deployments. 

\section{Conclusion}
\label{sec:conclusion}
\ac{RFF} has evolved from a narrow radar-specific technique identification into a broad, active research area at the intersection of signal processing, embedded systems, wireless security, and machine learning. This paper traced back that evolution across more than three decades, showing how the field moved through a sequence of major conceptual and methodological transitions. What began with the search for distinctive transient signatures gradually expanded into the study of steady-state impairments, then evolved into feature-engineered machine learning pipelines, and finally shifted toward deep representation learning from signal data. Each of these stages contributed important insights, but each also exposed limitations that shaped the next phase of research.

A central finding emerging from this historical analysis is that progress in \ac{RFF} has never been defined solely by improvements in classification accuracy. Early transient-based approaches demonstrated the physical plausibility of identifying transmitters from hardware-dependent imperfections, but short duration and the difficulty of data collection during the turn-on event constrained them. The later shift to digital communications made \ac{RFF} more relevant to practical security settings by focusing on features observable in ordinary transmissions. Subsequently, the \ac{ML} period standardized the fingerprinting workflow, making device authentication more systematic, while the \ac{DL} period reduced dependence on handcrafted features and enabled substantial performance gains in controlled settings. However, this progression also revealed that high accuracy does not necessarily imply the learning of stable, transmitter-specific representations 

The most important unresolved challenge remains the gap between laboratory performance and real-world deployment. Across the literature, \ac{RFF} systems continue to be affected by channel dependence, receiver sensitivity, session variability, limited dataset realism, and restricted generalization across environments, devices, and collection conditions. In addition, the rise of adversarial machine learning and active signal manipulation has shown that security-oriented evaluation must go far beyond benign closed-set classification. These issues are not marginal; they determine whether \ac{RFF} can serve as a trustworthy physical-layer security primitive in operational environments.

For this reason, the present and future of \ac{RFF} should be understood as a shift from accuracy-driven research to credibility-driven research. The next stage of the field will depend on more realistic datasets, reproducible benchmarks, open-set and open-world formulations, class-incremental learning, and robustness against both environmental variation and deliberate attacks. In parallel, we highlight the need for better isolation of transmitter-specific information from channel and receiver artifacts so that reported performance reflects genuine device identity rather than incidental measurement conditions. In this sense, the core question is no longer whether radios can be distinguished under favorable settings, but whether such distinctions remain stable, interpretable, and defensible when the assumptions of the laboratory are relaxed toward real-world deployment.

The path forward lies not simply in building larger (or more complex) models or achieving higher performance, but in developing evaluation practices and system designs that support reliable deployment in real, possibly contested, wireless environments. If these challenges are addressed carefully, \ac{RFF} can evolve into a practical and robust tool for next-generation wireless identification, monitoring, and security.


\bibliographystyle{IEEEtran}
\bibliography{main}

@inproceedings{toonstra_radio_1996_2,
	title = {A radio transmitter fingerprinting system {ODO}-1},
	volume = {1},
	doi = {10.1109/CCECE.1996.548038},
	booktitle = {Proceedings of 1996 {Canadian} {Conference} on {Electrical} and {Computer} {Engineering}},
	author = {Toonstra, J. and Kinsner, W.},
	year = {1996},
	pages = {60--63 vol.1},
}

@inproceedings{ureten_bayesian_1999_4,
	title = {Bayesian detection of radio transmitter turn-on transients},
	booktitle = {Proc. {NSIP99}},
	author = {Ureten, O. and Serinken, N.},
	year = {1999},
	pages = {830--834},
}

@inproceedings{shaw_multifractal_1997_5,
	title = {Multifractal modelling of radio transmitter transients for classification},
	doi = {10.1109/WESCAN.1997.627159},
	booktitle = {{IEEE} {WESCANEX} 97 {Communications}, {Power} and {Computing}. {Conference} {Proceedings}},
	author = {Shaw, D. and Kinsner, W.},
	year = {1997},
	pages = {306--312},
}

@inproceedings{langley_specific_1993_6,
	title = {Specific emitter identification ({SEI}) and classical parameter fusion technology},
	
	urldate = {2025-02-17},
	booktitle = {Proceedings of {WESCON}'93},
	publisher = {IEEE},
	author = {Langley, Lawrence E.},
	year = {1993},
	pages = {377--381},
}

@inproceedings{toonstra_transient_1995_7,
	title = {Transient analysis and genetic algorithms for classification},
	volume = {2},
	doi = {10.1109/WESCAN.1995.494069},
	booktitle = {{IEEE} {WESCANEX} 95. {Communications}, {Power}, and {Computing}. {Conference} {Proceedings}},
	author = {Toonstra, J. and Kinsner, W.},
	year = {1995},
	pages = {432--437 vol.2},
}

@inproceedings{brik_wireless_2008_14,
	address = {San Francisco California USA},
	title = {Wireless device identification with radiometric signatures},
	isbn = {978-1-60558-096-8},
	
	doi = {10.1145/1409944.1409959},
	language = {en},
	urldate = {2025-06-01},
	booktitle = {Proceedings of the 14th {ACM} international conference on {Mobile} computing and networking},
	publisher = {ACM},
	author = {Brik, Vladimir and Banerjee, Suman and Gruteser, Marco and Oh, Sangho},
	month = sep,
	year = {2008},
	pages = {116--127},
}

@inproceedings{li_combining_2009_20,
	title = {Combining {Multiple} {SVM} {Classifiers} for {Radar} {Emitter} {Recognition}},
	volume = {1},
	doi = {10.1109/FSKD.2009.623},
	booktitle = {2009 {Sixth} {International} {Conference} on {Fuzzy} {Systems} and {Knowledge} {Discovery}},
	author = {Li, Lin and Ji, Hongbing},
	year = {2009},
	pages = {140--144},
}

@inproceedings{yu_feature_2009_22,
	title = {Feature {Extraction} of {Radar} {Emitter} {Harmonic} {Power} {Constraint} {Based} on {Nonlinear} {Characters} of the {Amplifier}},
	doi = {10.1109/CISP.2009.5303962},
	booktitle = {2009 2nd {International} {Congress} on {Image} and {Signal} {Processing}},
	author = {Yu, Zhibin and Chen, Chunxia and Jin, Weidong and Zhang, Gexiang},
	year = {2009},
	pages = {1--4},
}

@incollection{paillier_rf-dna_2007_23,
	address = {Berlin, Heidelberg},
	title = {{RF}-{DNA}: {Radio}-{Frequency} {Certificates} of {Authenticity}},
	volume = {4727},
	isbn = {978-3-540-74734-5 978-3-540-74735-2},
	shorttitle = {{RF}-{DNA}},
	
	language = {en},
	urldate = {2025-06-01},
	booktitle = {Cryptographic {Hardware} and {Embedded} {Systems} - {CHES} 2007},
	publisher = {Springer Berlin Heidelberg},
	author = {DeJean, Gerald and Kirovski, Darko},
	editor = {Paillier, Pascal and Verbauwhede, Ingrid},
	year = {2007},
	doi = {10.1007/978-3-540-74735-2_24},
	note = {ISSN: 0302-9743, 1611-3349
Series Title: Lecture Notes in Computer Science},
	pages = {346--363},
}

@inproceedings{cobb_physical_2010_28,
	title = {Physical layer identification of embedded devices using {RF}-{DNA} fingerprinting},
	
	urldate = {2025-06-01},
	booktitle = {2010-{Milcom} 2010 {Military} {Communications} {Conference}},
	publisher = {IEEE},
	author = {Cobb, William E. and Garcia, Eric W. and Temple, Michael A. and Baldwin, Rusty O. and Kim, Yong C.},
	year = {2010},
	pages = {2168--2173},
}

@inproceedings{kennedy_radio_2008_29,
	title = {Radio transmitter fingerprinting: {A} steady state frequency domain approach},
	shorttitle = {Radio transmitter fingerprinting},
	
	urldate = {2025-06-01},
	booktitle = {2008 {IEEE} 68th {Vehicular} {Technology} {Conference}},
	publisher = {IEEE},
	author = {Kennedy, Irwin O. and Scanlon, Patricia and Mullany, Francis J. and Buddhikot, Milind M. and Nolan, Keith E. and Rondeau, Thomas W.},
	year = {2008},
	pages = {1--5},
}

@inproceedings{candore_robust_2009_31,
	title = {Robust stable radiometric fingerprinting for wireless devices},
	doi = {10.1109/HST.2009.5224969},
	booktitle = {2009 {IEEE} {International} {Workshop} on {Hardware}-{Oriented} {Security} and {Trust}},
	author = {Candore, Andrea and Kocabas, Ovunc and Koushanfar, Farinaz},
	year = {2009},
	pages = {43--49},
}

@article{ureten_wireless_2007_39,
	title = {Wireless security through {RF} fingerprinting},
	volume = {32},
	doi = {10.1109/CJECE.2007.364330},
	number = {1},
	journal = {Canadian Journal of Electrical and Computer Engineering},
	author = {Ureten, Oktay and Serinken, Nur},
	year = {2007},
	pages = {27--33},
}

@article{chapre_csi_2015_42,
  title={CSI-MIMO: An efficient Wi-Fi fingerprinting using channel state information with MIMO},
  author={Chapre, Yogita and Ignjatovic, Aleksandar and Seneviratne, Aruna and Jha, Sanjay},
  journal={Pervasive and Mobile Computing},
  volume={23},
  pages={89--103},
  year={2015},
  publisher={Elsevier}
}

@inproceedings{rehman_analysis_2012_45,
	title = {Analysis of receiver front end on the performance of {RF} fingerprinting},
	
	urldate = {2025-06-01},
	booktitle = {2012 {IEEE} 23rd {International} {Symposium} on {Personal}, {Indoor} and {Mobile} {Radio} {Communications}-({PIMRC})},
	publisher = {IEEE},
	author = {Rehman, Saeed Ur and Sowerby, Kevin and Coghill, Colin},
	year = {2012},
	pages = {2494--2499},
}

@inproceedings{patel_comparison_2014_47,
	title = {Comparison of high-end and low-end receivers for {RF}-{DNA} fingerprinting},
	
	urldate = {2025-06-01},
	booktitle = {2014 {IEEE} {Military} {Communications} {Conference}},
	publisher = {IEEE},
	author = {Patel, Hiren and Temple, Michael A. and Ramsey, Benjamin W.},
	year = {2014},
	pages = {24--29},
}

@article{reising_authorized_2015_49,
	title = {Authorized and rogue device discrimination using dimensionally reduced {RF}-{DNA} fingerprints},
	volume = {10},
	
	number = {6},
	urldate = {2025-06-01},
	journal = {IEEE Transactions on Information Forensics and Security},
	author = {Reising, Donald R. and Temple, Michael A. and Jackson, Julie A.},
	year = {2015},
	note = {Publisher: IEEE},
	pages = {1180--1192},
}

@inproceedings{revadigar_dlink_2015_50,
	title = {{DLINK}: {Dual} link based radio frequency fingerprinting for wearable devices},
	shorttitle = {{DLINK}},
	
	urldate = {2025-06-01},
	booktitle = {2015 {IEEE} 40th {Conference} on {Local} {Computer} {Networks} ({LCN})},
	publisher = {IEEE},
	author = {Revadigar, Girish and Javali, Chitra and Hu, Wen and Jha, Sanjay},
	year = {2015},
	pages = {329--337},
}

@article{polak_identification_2015_53,
	title = {Identification of {Wireless} {Devices} of {Users} {Who} {Actively} {Fake} {Their} {RF} {Fingerprints} {With} {Artificial} {Data} {Distortion}},
	volume = {14},
	doi = {10.1109/TWC.2015.2443794},
	number = {11},
	journal = {IEEE Transactions on Wireless Communications},
	author = {Polak, Adam C. and Goeckel, Dennis L.},
	year = {2015},
	pages = {5889--5899},
}

@article{liu_nonlinearity_2011_56,
	title = {Nonlinearity {Estimation} for {Specific} {Emitter} {Identification} in {Multipath} {Channels}},
	volume = {6},
	doi = {10.1109/TIFS.2011.2134848},
	number = {3},
	journal = {IEEE Transactions on Information Forensics and Security},
	author = {Liu, Ming-Wei and Doherty, John F.},
	year = {2011},
	pages = {1076--1085},
}

@article{patel_improving_2014_57,
	title = {Improving {ZigBee} device network authentication using ensemble decision tree classifiers with radio frequency distinct native attribute fingerprinting},
	volume = {64},
	
	number = {1},
	urldate = {2025-06-01},
	journal = {IEEE transactions on reliability},
	author = {Patel, Hiren J. and Temple, Michael A. and Baldwin, Rusty O.},
	year = {2014},
	note = {Publisher: IEEE},
	pages = {221--233},
}

@article{polak_wireless_2015_67,
	title = {Wireless {Device} {Identification} {Based} on {RF} {Oscillator} {Imperfections}},
	volume = {10},
	doi = {10.1109/TIFS.2015.2464778},
	number = {12},
	journal = {IEEE Transactions on Information Forensics and Security},
	author = {Polak, Adam C. and Goeckel, Dennis L.},
	year = {2015},
	pages = {2492--2501},
}

@inproceedings{wheeler_assessment_2017_72,
	title = {Assessment of the impact of {CFO} on {RF}-{DNA} fingerprint classification performance},
	
	urldate = {2025-06-01},
	booktitle = {2017 {International} {Conference} on {Computing}, {Networking} and {Communications} ({ICNC})},
	publisher = {IEEE},
	author = {Wheeler, Charles G. and Reising, Donald R.},
	year = {2017},
	pages = {110--114},
}

@inproceedings{vo-huu_fingerprinting_2016_74,
	address = {Darmstadt Germany},
	title = {Fingerprinting {Wi}-{Fi} {Devices} {Using} {Software} {Defined} {Radios}},
	isbn = {978-1-4503-4270-4},
	
	doi = {10.1145/2939918.2939936},
	language = {en},
	urldate = {2025-06-01},
	booktitle = {Proceedings of the 9th {ACM} {Conference} on {Security} \& {Privacy} in {Wireless} and {Mobile} {Networks}},
	publisher = {ACM},
	author = {Vo-Huu, Tien Dang and Vo-Huu, Triet Dang and Noubir, Guevara},
	month = jul,
	year = {2016},
	pages = {3--14},
}

@inproceedings{zhuang_fbsleuth_2018_84,
	address = {Incheon Republic of Korea},
	title = {{FBSleuth}: {Fake} {Base} {Station} {Forensics} via {Radio} {Frequency} {Fingerprinting}},
	isbn = {978-1-4503-5576-6},
	shorttitle = {{FBSleuth}},
	
	doi = {10.1145/3196494.3196521},
	language = {en},
	urldate = {2025-06-01},
	booktitle = {Proceedings of the 2018 on {Asia} {Conference} on {Computer} and {Communications} {Security}},
	publisher = {ACM},
	author = {Zhuang, Zhou and Ji, Xiaoyu and Zhang, Taimin and Zhang, Juchuan and Xu, Wenyuan and Li, Zhenhua and Liu, Yunhao},
	month = may,
	year = {2018},
	pages = {261--272},
}

@article{merchant_deep_2018_85,
	title = {Deep learning for {RF} device fingerprinting in cognitive communication networks},
	volume = {12},
	
	number = {1},
	urldate = {2025-06-01},
	journal = {IEEE journal of selected topics in signal processing},
	author = {Merchant, Kevin and Revay, Shauna and Stantchev, George and Nousain, Bryan},
	year = {2018},
	note = {Publisher: IEEE},
	pages = {160--167},
}

@article{peng_design_2018_86,
	title = {Design of a hybrid {RF} fingerprint extraction and device classification scheme},
	volume = {6},
	
	number = {1},
	urldate = {2025-06-01},
	journal = {IEEE internet of things journal},
	author = {Peng, Linning and Hu, Aiqun and Zhang, Junqing and Jiang, Yu and Yu, Jiabao and Yan, Yan},
	year = {2018},
	note = {Publisher: IEEE},
	pages = {349--360},
}

@article{chen_lightweight_2018_87,
	title = {Lightweight one‐time password authentication scheme based on radio‐frequency fingerprinting},
	volume = {12},
	issn = {1751-8628, 1751-8636},
	
	doi = {10.1049/iet-com.2018.0023},
	language = {en},
	number = {12},
	urldate = {2025-06-01},
	journal = {IET Communications},
	author = {Chen, Yi and Wen, Hong and Song, Huanhuan and Chen, Songlin and Xie, Feiyi and Yang, Qing and Hu, Lin},
	month = jul,
	year = {2018},
	pages = {1477--1484},
}

@inproceedings{jafari_iot_2018_88,
  title={IoT devices fingerprinting using deep learning},
  author={Jafari, Hossein and Omotere, Oluwaseyi and Adesina, Damilola and Wu, Hsiang-Huang and Qian, Lijun},
  booktitle={MILCOM 2018-2018 IEEE Military Communications Conference (MILCOM)},
  pages={1--9},
  year={2018},
  organization={IEEE}
}

@article{youssef_machine_2018_90,
  title={Machine learning approach to RF transmitter identification},
  author={Youssef, Khalid and Bouchard, Louis and Haigh, Karen and Silovsky, Jan and Thapa, Bishal and Vander Valk, Chris},
  journal={IEEE Journal of Radio Frequency Identification},
  volume={2},
  number={4},
  pages={197--205},
  year={2018},
  publisher={IEEE}
}

@book{riyaz_radio_2018_91,
  title={Radio Fingerprinting Using Convolutional Neural Networks},
  author={Riyaz, Shamnaz Mohammed},
  year={2018},
  publisher={Northeastern University}
}

@article{ding_specific_2018_95,
  title={Specific emitter identification via convolutional neural networks},
  author={Ding, Lida and Wang, Shilian and Wang, Fanggang and Zhang, Wei},
  journal={IEEE communications letters},
  volume={22},
  number={12},
  pages={2591--2594},
  year={2018},
  publisher={IEEE}
}

@inproceedings{restuccia_deepradioid_2019_97,
	address = {Catania Italy},
	title = {{DeepRadioID}: {Real}-{Time} {Channel}-{Resilient} {Optimization} of {Deep} {Learning}-based {Radio} {Fingerprinting} {Algorithms}},
	isbn = {978-1-4503-6764-6},
	shorttitle = {{DeepRadioID}},
	
	doi = {10.1145/3323679.3326503},
	language = {en},
	urldate = {2025-06-01},
	booktitle = {Proceedings of the {Twentieth} {ACM} {International} {Symposium} on {Mobile} {Ad} {Hoc} {Networking} and {Computing}},
	publisher = {ACM},
	author = {Restuccia, Francesco and D'Oro, Salvatore and Al-Shawabka, Amani and Belgiovine, Mauro and Angioloni, Luca and Ioannidis, Stratis and Chowdhury, Kaushik and Melodia, Tommaso},
	month = jul,
	year = {2019},
	pages = {51--60},
}

@article{peng_deep_2019_99,
	title = {Deep learning based {RF} fingerprint identification using differential constellation trace figure},
	volume = {69},
	
	number = {1},
	urldate = {2025-06-01},
	journal = {IEEE Transactions on Vehicular Technology},
	author = {Peng, Linning and Zhang, Junqing and Liu, Ming and Hu, Aiqun},
	year = {2019},
	note = {Publisher: IEEE},
	pages = {1091--1095},
}

@inproceedings{gritsenko_finding_2019_100,
	title = {Finding a ‘new’needle in the haystack: {Unseen} radio detection in large populations using deep learning},
	shorttitle = {Finding a ‘new’needle in the haystack},
	
	urldate = {2025-06-01},
	booktitle = {2019 {IEEE} international symposium on dynamic spectrum access networks ({DySPAN})},
	publisher = {IEEE},
	author = {Gritsenko, Andrey and Wang, Zifeng and Jian, Tong and Dy, Jennifer and Chowdhury, Kaushik and Ioannidis, Stratis},
	year = {2019},
	pages = {1--10},
}

@article{sankhe_no_2019_102,
	title = {No radio left behind: {Radio} fingerprinting through deep learning of physical-layer hardware impairments},
	volume = {6},
	shorttitle = {No radio left behind},
	
	number = {1},
	urldate = {2025-06-01},
	journal = {IEEE Transactions on Cognitive Communications and Networking},
	author = {Sankhe, Kunal and Belgiovine, Mauro and Zhou, Fan and Angioloni, Luca and Restuccia, Frank and D’Oro, Salvatore and Melodia, Tommaso and Ioannidis, Stratis and Chowdhury, Kaushik},
	year = {2019},
	note = {Publisher: IEEE},
	pages = {165--178},
}

@inproceedings{fadul_rf-dna_2019_109,
	title = {{RF}-{DNA} fingerprint classification of {OFDM} signals using a {Rayleigh} fading channel model},
	
	urldate = {2025-06-01},
	booktitle = {2019 {IEEE} {Wireless} {Communications} and {Networking} {Conference} ({WCNC})},
	publisher = {IEEE},
	author = {Fadul, Mohamed KM and Reising, Donald R. and Loveless, T. Daniel and Ofoli, Abdul R.},
	year = {2019},
	pages = {1--7},
}

@article{jian_deep_2020_120,
	title = {Deep learning for {RF} fingerprinting: {A} massive experimental study},
	volume = {3},
	shorttitle = {Deep learning for {RF} fingerprinting},
	
	number = {1},
	urldate = {2025-06-01},
	journal = {IEEE Internet of Things Magazine},
	author = {Jian, Tong and Rendon, Bruno Costa and Ojuba, Emmanuel and Soltani, Nasim and Wang, Zifeng and Sankhe, Kunal and Gritsenko, Andrey and Dy, Jennifer and Chowdhury, Kaushik and Ioannidis, Stratis},
	year = {2020},
	note = {Publisher: IEEE},
	pages = {50--57},
}

@inproceedings{al-shawabka_exposing_2020_123,
	title = {Exposing the fingerprint: {Dissecting} the impact of the wireless channel on radio fingerprinting},
	shorttitle = {Exposing the fingerprint},
	
	urldate = {2025-06-01},
	booktitle = {{IEEE} {INFOCOM} 2020-{IEEE} {Conference} on {Computer} {Communications}},
	publisher = {IEEE},
	author = {Al-Shawabka, Amani and Restuccia, Francesco and D’Oro, Salvatore and Jian, Tong and Rendon, Bruno Costa and Soltani, Nasim and Dy, Jennifer and Ioannidis, Stratis and Chowdhury, Kaushik and Melodia, Tommaso},
	year = {2020},
	pages = {646--655},
}

@article{wang_efficient_2021_138,
	title = {An {Efficient} {Specific} {Emitter} {Identification} {Method} {Based} on {Complex}-{Valued} {Neural} {Networks} and {Network} {Compression}},
	volume = {39},
	issn = {1558-0008},
	
	doi = {10.1109/JSAC.2021.3087243},
	number = {8},
	urldate = {2025-07-14},
	journal = {IEEE Journal on Selected Areas in Communications},
	author = {Wang, Yu and Gui, Guan and Gacanin, Haris and Ohtsuki, Tomoaki and Dobre, Octavia A. and Poor, H. Vincent},
	month = aug,
	year = {2021},
	pages = {2305--2317},
}

@article{shen_radio_2021_139,
	title = {Radio {Frequency} {Fingerprint} {Identification} for {LoRa} {Using} {Deep} {Learning}},
	volume = {39},
	issn = {1558-0008},
	
	doi = {10.1109/JSAC.2021.3087250},
	number = {8},
	urldate = {2025-06-25},
	journal = {IEEE Journal on Selected Areas in Communications},
	author = {Shen, Guanxiong and Zhang, Junqing and Marshall, Alan and Peng, Linning and Wang, Xianbin},
	month = aug,
	year = {2021},
	pages = {2604--2616},
}

@article{huang_deep_2022_149,
	title = {Deep {Learning} of {Radio} {Frequency} {Fingerprints} from {Limited} {Samples} by {Masked} {Autoencoding}},
	issn = {2162-2345},
	
	doi = {10.1109/LWC.2022.3184674},
	urldate = {2025-07-14},
	journal = {IEEE Wireless Communications Letters},
	author = {Huang, Keju and Yang, Jun’an and Liu, Hui and Hu, Pengjiang},
	year = {2022},
	pages = {1--1},
}

@inproceedings{tyler_analysis_2022_151,
	title = {An {Analysis} of {Signal} {Energy} {Impacts} and {Threats} to {Deep} {Learning} {Based} {SEI}},
	doi = {10.1109/ICC45855.2022.9838884},
	booktitle = {{ICC} 2022 - {IEEE} {International} {Conference} on {Communications}},
	author = {Tyler, Joshua H. and Fadul, Mohamed M. K. and Reising, Donald R. and Kandah, Farah I.},
	year = {2022},
	pages = {2077--2083},
}

@article{wang_few-shot_2022_157,
	title = {Few-{Shot} {Specific} {Emitter} {Identification} via {Deep} {Metric} {Ensemble} {Learning}},
	volume = {9},
	issn = {2327-4662},
	
	doi = {10.1109/JIOT.2022.3194967},
	number = {24},
	urldate = {2025-07-03},
	journal = {IEEE Internet of Things Journal},
	author = {Wang, Yu and Gui, Guan and Lin, Yun and Wu, Hsiao-Chun and Yuen, Chau and Adachi, Fumiyuki},
	month = dec,
	year = {2022},
	pages = {24980--24994},
}

@article{sun_robustness_2022_161,
	title = {Robustness of {Deep} {Learning}-{Based} {Specific} {Emitter} {Identification} under {Adversarial} {Attacks}},
	volume = {14},
	copyright = {http://creativecommons.org/licenses/by/3.0/},
	issn = {2072-4292},
	
	doi = {10.3390/rs14194996},
	language = {en},
	number = {19},
	urldate = {2025-07-03},
	journal = {Remote Sensing},
	author = {Sun, Liting and Ke, Da and Wang, Xiang and Huang, Zhitao and Huang, Kaizhu},
	month = jan,
	year = {2022},
	note = {Number: 19
Publisher: Multidisciplinary Digital Publishing Institute},
	pages = {4996},
}

@article{shen_towards_2022_162,
	title = {Towards {Scalable} and {Channel}-{Robust} {Radio} {Frequency} {Fingerprint} {Identification} for {LoRa}},
	volume = {17},
	issn = {1556-6021},
	
	doi = {10.1109/TIFS.2022.3152404},
	urldate = {2025-06-25},
	journal = {IEEE Transactions on Information Forensics and Security},
	author = {Shen, Guanxiong and Zhang, Junqing and Marshall, Alan and Cavallaro, Joseph R.},
	year = {2022},
	pages = {774--787},
}

@article{yang_specific_2022_163,
	title = {Specific {Emitter} {Identification} {With} {Limited} {Samples}: {A} {Model}-{Agnostic} {Meta}-{Learning} {Approach}},
	volume = {26},
	issn = {1558-2558},
	shorttitle = {Specific {Emitter} {Identification} {With} {Limited} {Samples}},
	
	doi = {10.1109/LCOMM.2021.3110775},
	number = {2},
	urldate = {2025-07-03},
	journal = {IEEE Communications Letters},
	author = {Yang, Ning and Zhang, Bangning and Ding, Guoru and Wei, Yimin and Wei, Guofeng and Wang, Jian and Guo, Daoxing},
	month = feb,
	year = {2022},
	pages = {345--349},
}

@article{oligeri_past-ai_2023_180,
	title = {{PAST}-{AI}: {Physical}-{Layer} {Authentication} of {Satellite} {Transmitters} via {Deep} {Learning}},
	volume = {18},
	issn = {1556-6021},
	shorttitle = {{PAST}-{AI}},
	
	doi = {10.1109/TIFS.2022.3219287},
	urldate = {2025-06-01},
	journal = {IEEE Transactions on Information Forensics and Security},
	author = {Oligeri, Gabriele and Sciancalepore, Savio and Raponi, Simone and Pietro, Roberto Di},
	year = {2023},
	pages = {274--289},
}

@inproceedings{cui_performance_2023_181,
	title = {Performance {Evaluation} of {RF} {Fingerprinting} via {Deep} {Learning} for {Satellite} {Network} {Security}},
	doi = {10.1109/iThings-GreenCom-CPSCom-SmartData-Cybermatics60724.2023.00061},
	booktitle = {2023 {IEEE} {International} {Conferences} on {Internet} of {Things} ({iThings}) and {IEEE} {Green} {Computing} \& {Communications} ({GreenCom}) and {IEEE} {Cyber}, {Physical} \& {Social} {Computing} ({CPSCom}) and {IEEE} {Smart} {Data} ({SmartData}) and {IEEE} {Congress} on {Cybermatics} ({Cybermatics})},
	author = {Cui, Tianshu and Yu, Yang and Liu, Yang and Lu, Xinyu and Rao, Jiacheng and Zhang, You},
	month = dec,
	year = {2023},
	note = {ISSN: 2836-3701},
	pages = {246--251},
}

@inproceedings{reising_rf_2023_188,
	title = {{RF} {Fingerprint}-based {Identity} {Verification} in the {Presence} of an {SEI} {Mimicking} {Adversary}},
	doi = {10.1109/WiMob58348.2023.10187867},
	booktitle = {2023 19th {International} {Conference} on {Wireless} and {Mobile} {Computing}, {Networking} and {Communications} ({WiMob})},
	author = {Reising, Donald R. and Tyler, Joshua H. and Fadul, Mohamed K. M. and Hilling, Matthew R. and Loveless, T. Daniel},
	year = {2023},
	pages = {438--444},
}

@article{papangelo_adversarial_2024_199,
	title = {Adversarial {Machine} {Learning} for {Image}-{Based} {Radio} {Frequency} {Fingerprinting}: {Attacks} and {Defenses}},
	volume = {62},
	issn = {1558-1896},
	doi = {10.1109/MCOM.001.2300464},
	number = {11},
	journal = {IEEE Communications Magazine},
	author = {Papangelo, Lorenzo and Pistilli, Maurizio and Sciancalepore, Savio and Oligeri, Gabriele and Piro, Giuseppe and Boggia, Gennaro},
	month = nov,
	year = {2024},
	pages = {108--113},
}

@inproceedings{oligeri_fadeprint_2024_210,
	title = {{FadePrint} - {Satellite} {Spoofing} {Detection} via {Fading} {Fingerprinting}},
	doi = {10.1109/CCNC51664.2024.10454707},
	booktitle = {2024 {IEEE} 21st {Consumer} {Communications} \& {Networking} {Conference} ({CCNC})},
	author = {Oligeri, Gabriele and Sciancalepore, Savio and Sadighian, Alireza},
	month = jan,
	year = {2024},
	note = {ISSN: 2331-9860},
	pages = {827--830},
}

@inproceedings{givehchian_practical_2024_219,
	title = {Practical {Obfuscation} of {BLE} {Physical}-{Layer} {Fingerprints} on {Mobile} {Devices}},
	
	doi = {10.1109/SP54263.2024.00073},
	urldate = {2025-03-16},
	booktitle = {2024 {IEEE} {Symposium} on {Security} and {Privacy} ({SP})},
	author = {Givehchian, Hadi and Bhaskar, Nishant and Redding, Alexander and Zhao, Han and Schulman, Aaron and Bharadia, Dinesh},
	month = may,
	year = {2024},
	note = {ISSN: 2375-1207},
	pages = {2867--2885},
}

@inproceedings{al-hazbi_radio_2024_221,
	title = {Radio {Frequency} {Fingerprinting} via {Deep} {Learning}: {Challenges} and {Opportunities}},
	doi = {10.1109/IWCMC61514.2024.10592579},
	booktitle = {2024 {International} {Wireless} {Communications} and {Mobile} {Computing} ({IWCMC})},
	author = {Al-Hazbi, Saeif and Hussain, Ahmed and Sciancalepore, Savio and Oligeri, Gabriele and Papadimitratos, Panos},
	month = may,
	year = {2024},
	note = {ISSN: 2376-6506},
	pages = {0824--0829},
}

@article{muller_sensitivity_2024_224,
	title = {Sensitivity {Analysis} of {RFML} {Applications}},
	volume = {12},
	doi = {10.1109/ACCESS.2024.3409471},
	journal = {IEEE Access},
	author = {Muller, Braeden P. and Wong, Lauren J. and Michaels, Alan J.},
	year = {2024},
	pages = {80327--80344},
}

@article{han_open-world_2025_235,
	title = {Open-world {Radio} {Frequency} {Fingerprint} {Identification} via {Augmented} {Semi}-supervised {Learning}},
	volume = {39},
	copyright = {Copyright (c) 2025 Association for the Advancement of Artificial Intelligence},
	issn = {2374-3468},
	
	doi = {10.1609/aaai.v39i1.32003},
	language = {en},
	number = {1},
	urldate = {2025-04-13},
	journal = {Proceedings of the AAAI Conference on Artificial Intelligence},
	author = {Han, Zehua and Xiao, Jing and Zhao, Qirui and Cui, Zhexuan and Wang, Yufeng and Zhang, Duona and Ding, Wenrui},
	month = apr,
	year = {2025},
	note = {Number: 1},
	pages = {264--272},
}

@inproceedings{fu_channel-robust_2025_241,
	title = {Channel-{Robust} {Few}-{Shot} {Specific} {Emitter} {Identification} {Using} {Meta}-{Feature} {Augmentation}},
	
	doi = {10.1109/WCNC61545.2025.10978187},
	urldate = {2025-05-15},
	booktitle = {2025 {IEEE} {Wireless} {Communications} and {Networking} {Conference} ({WCNC})},
	author = {Fu, Xue and Meneghello, Francesca and Wang, Yu and Ohtsuki, Tomoaki and Yuen, Chau and Gui, Guan and Sari, Hikmet},
	month = mar,
	year = {2025},
	note = {ISSN: 1558-2612},
	pages = {1--6},
}

@article{li_fscil-sei_2025_251,
	title = {{FSCIL}-{SEI}: {Few}-{Shot} {Class}-{Incremental} {Learning} {Approach} for {Specific} {Emitter} {Identification}},
	volume = {74},
	issn = {1557-9662},
	doi = {10.1109/TIM.2025.3529056},
	journal = {IEEE Transactions on Instrumentation and Measurement},
	author = {Li, Dingzhao and Chen, Xiaowei and Hong, Shaohua and Qi, Jie and Sun, Haixin},
	year = {2025},
	pages = {1--14},
}

@inproceedings{baldini_mitigation_2025_259,
	title = {Mitigation of {Poisoning} {Attacks} in {Radio} {Frequency} {Fingerprinting} ({RFF}) with {Outlier} {Detection} {Algorithms}},
	
	doi = {10.1109/MeditCom64437.2025.11104347},
	urldate = {2025-09-07},
	booktitle = {2025 {IEEE} {International} {Mediterranean} {Conference} on {Communications} and {Networking} ({MeditCom})},
	author = {Baldini, Gianmarco},
	month = jul,
	year = {2025},
	pages = {1--6},
}

@inproceedings{hou_prototype-guided_2025_261,
	title = {Prototype-{Guided} {Anti}-{Jamming} {Framework} {Against} {White}-{Box} {Adversarial} {Attack} in {Radio}-{Frequency} {Fingerprint} {Identification}},
	
	doi = {10.1109/ICCC65529.2025.11149102},
	urldate = {2025-09-18},
	booktitle = {2025 {IEEE}/{CIC} {International} {Conference} on {Communications} in {China} ({ICCC})},
	author = {Hou, Yuzhen and Zhang, Tiantian and Cheng, Nan and Sun, Ruijin and Li, Changle},
	month = aug,
	year = {2025},
	note = {ISSN: 2377-8644},
	pages = {1--6},
}

@article{jing_dynamic_2025_264,
	title = {Dynamic {Boundary} {Adversarial} {Model} for {Open} {Set} {Radar}-{Specific} {Emitter} {Identification}},
	issn = {1557-9603},
	
	doi = {10.1109/TAES.2025.3581888},
	urldate = {2025-09-28},
	journal = {IEEE Transactions on Aerospace and Electronic Systems},
	author = {Jing, Zehuan and Li, Peng and Zhou, Xu and Yan, Erxing and Chen, Yingchao and Li, Jingyi and Wang, Zhao},
	year = {2025},
	pages = {1--25},
}

@article{bothereau_why_2025_275,
	title = {Why {RF} {Fingerprinting} {Needs} {Better} {Data}, {Not} {Bigger} {Models}},
	issn = {2169-3536},
	
	doi = {10.1109/ACCESS.2025.3614459},
	urldate = {2025-09-29},
	journal = {IEEE Access},
	author = {Bothereau, Emma and Gerzaguet, Robin and Gautier, Matthieu and Chillet, Alice and Berder, Olivier},
	year = {2025},
	pages = {1--1},
}

@article{jagannath_comprehensive_2022,
	title = {A comprehensive survey on radio frequency ({RF}) fingerprinting: {Traditional} approaches, deep learning, and open challenges},
	volume = {219},
	issn = {1389-1286},
	shorttitle = {A comprehensive survey on radio frequency ({RF}) fingerprinting},
	doi = {10.1016/j.comnet.2022.109455},
	urldate = {2025-06-25},
	journal = {Computer Networks},
	author = {Jagannath, Anu and Jagannath, Jithin and Kumar, Prem Sagar Pattanshetty Vasanth},
	month = dec,
	year = {2022},
	pages = {109455},
}

@inproceedings{sagduyu_adversarial_2023,
	title = {Adversarial {Attacks} on {LoRa} {Device} {Identification} and {Rogue} {Signal} {Detection} with {Deep} {Learning}},
	issn = {2155-7586},
	doi = {10.1109/MILCOM58377.2023.10356224},
	urldate = {2026-02-04},
	booktitle = {{MILCOM} 2023 - 2023 {IEEE} {Military} {Communications} {Conference} ({MILCOM})},
	author = {Sagduyu, Yalin E. and Erpek, Tugba},
	month = oct,
	year = {2023},
	note = {ISSN: 2155-7586},
	pages = {385--390},
}

@article{huang_radio_2024,
	title = {Radio frequency fingerprint extraction and authentication towards open set in noisy channels},
	volume = {146},
	issn = {1051-2004},
	doi = {10.1016/j.dsp.2023.104363},
	urldate = {2026-02-04},
	journal = {Digital Signal Processing},
	author = {Huang, Renhui and Peng, Xinyong and Chai, Zhi and Li, Mingye and Ren, Jiawei and Yang, Xuelin},
	month = mar,
	year = {2024},
	pages = {104363},
}

@article{zhang_real-world_2024,
	title = {Real-{World} {Aircraft} {Recognition} {Based} on {RF} {Fingerprinting} {With} {Few} {Labeled} {ADS}-{B} {Signals}},
	volume = {73},
	issn = {1939-9359},
	doi = {10.1109/TVT.2023.3314491},
	number = {2},
	urldate = {2026-02-04},
	journal = {IEEE Transactions on Vehicular Technology},
	author = {Zhang, Zechen and Li, Guyue and Shi, Jitong and Li, Haobo and Hu, Aiqun},
	month = feb,
	year = {2024},
	pages = {2866--2871},
}

@inproceedings{smailes_watch_2023,
	address = {Copenhagen Denmark},
	title = {Watch {This} {Space}: {Securing} {Satellite} {Communication} through {Resilient} {Transmitter} {Fingerprinting}},
	isbn = {979-8-4007-0050-7},
	shorttitle = {Watch {This} {Space}},
	doi = {10.1145/3576915.3623135},
	language = {en},
	urldate = {2025-10-12},
	booktitle = {Proceedings of the 2023 {ACM} {SIGSAC} {Conference} on {Computer} and {Communications} {Security}},
	publisher = {ACM},
	author = {Smailes, Joshua and Köhler, Sebastian and Birnbach, Simon and Strohmeier, Martin and Martinovic, Ivan},
	month = nov,
	year = {2023},
	pages = {608--621},
}

@article{tian_optimized_2024,
	title = {Optimized {Radio} {Frequency} {Footprint} {Identification} {Based} on {UAV} {Telemetry} {Radios}},
	volume = {24},
	copyright = {http://creativecommons.org/licenses/by/3.0/},
	issn = {1424-8220},
	doi = {10.3390/s24165099},
	language = {en},
	number = {16},
	urldate = {2026-02-05},
	journal = {Sensors},
	publisher = {publisher},
	author = {Tian, Yuan and Wen, Hong and Zhou, Jiaxin and Duan, Zhiqiang and Li, Tao},
	month = aug,
	year = {2024},
}

@ARTICLE{Zhang2025,
  author={Zhang, Junqing and Ardizzon, Francesco and Piana, Mattia and Shen, Guanxiong and Tomasin, Stefano},
  journal={IEEE Transactions on Information Forensics and Security}, 
  title={Physical Layer-Based Device Fingerprinting for Wireless Security: From Theory to Practice}, 
  year={2025},
  volume={20},
  number={},
  pages={5296-5325},
  doi={10.1109/TIFS.2025.3570118}}

@article{irfan_device_2025,
    title = {Device {Fingerprinting} in {Power} {Line} {Communications}},
    volume = {178},
    issn = {1570-8705},
    url = {https://www.sciencedirect.com/science/article/pii/S1570870525002033},
    doi = {10.1016/j.adhoc.2025.103955},
    urldate = {2025-09-21},
    journal = {Ad Hoc Networks},
    author = {Irfan, Muhammad and Hernandez Fernandez, Javier and Omri, Aymen and Sciancalepore, Savio and Oligeri, Gabriele},
    month = nov,
    year = {2025},
    pages = {103955},
}

@article{hanna_wisig_2022,
    title = {{WiSig}: {A} {Large}-{Scale} {WiFi} {Signal} {Dataset} for {Receiver} and {Channel} {Agnostic} {RF} {Fingerprinting}},
    volume = {10},
    issn = {2169-3536},
    shorttitle = {{WiSig}},
    url = {https://ieeexplore.ieee.org/document/9721895},
    doi = {10.1109/ACCESS.2022.3154790},
    urldate = {2026-05-20},
    journal = {IEEE Access},
    author = {Hanna, Samer and Karunaratne, Samurdhi and Cabric, Danijela},
    year = {2022},
    pages = {22808--22818},
}

@inproceedings{sankhe_oracle_2019,
    title = {{ORACLE}: {Optimized} {Radio} {clAssification} through {Convolutional} {neuraL} {nEtworks}},
    issn = {2641-9874},
    shorttitle = {{ORACLE}},
    url = {https://ieeexplore.ieee.org/document/8737463},
    doi = {10.1109/INFOCOM.2019.8737463},
    urldate = {2026-05-20},
    booktitle = {{IEEE} {INFOCOM} 2019 - {IEEE} {Conference} on {Computer} {Communications}},
    author = {Sankhe, Kunal and Belgiovine, Mauro and Zhou, Fan and Riyaz, Shamnaz and Ioannidis, Stratis and Chowdhury, Kaushik},
    month = apr,
    year = {2019},
    note = {ISSN: 2641-9874},
    pages = {370--378},
}

@misc{zhang_rapid_2026,
    title = {Rapid {LoRA} {Aggregation} for {Wireless} {Channel} {Adaptation} in {Open}-{Set} {Radio} {Frequency} {Fingerprinting}},
    url = {http://arxiv.org/abs/2604.12834},
    doi = {10.48550/arXiv.2604.12834},
    urldate = {2026-04-18},
    publisher = {arXiv},
    author = {Zhang, Mingxi and Xie, Renjie and Wang, Jincheng and Li, Guyue and Xu, Wei},
    month = apr,
    year = {2026},
    note = {arXiv:2604.12834 [eess]},
}

@inproceedings{dat_Alhazbi_2023,
author = {Saeif, Alhazbi and Savio, Sciancalepore and Gabriele, Oligeri},
title = {The Day-After-Tomorrow: On the Performance of Radio Fingerprinting over Time},
year = {2023},
isbn = {9798400708862},
publisher = {Association for Computing Machinery},
address = {New York, NY, USA},
url = {https://doi.org/10.1145/3627106.3627192},
doi = {10.1145/3627106.3627192},
booktitle = {Proceedings of the 39th Annual Computer Security Applications Conference},
pages = {439–450},
numpages = {12},
location = {Austin, TX, USA},
series = {ACSAC '23}
}

@INPROCEEDINGS{irfan_cps-sec_2025,
  author={Irfan, Muhammad and Oligeri, Gabriele and Sciancalepore, Savio},
  booktitle={2025 IEEE Conference on Communications and Network Security (CNS)}, 
  title={Shifting Signatures: The Ephemeral Nature of the Radio Fingerprint on the USRP X310}, 
  year={2025},
  volume={},
  number={},
  pages={1-6},
  doi={10.1109/CNS66487.2025.11194974}}

@article{temp_rff_gu2024,
title = {TEA-RFFI: Temperature adjusted radio frequency fingerprint-based smartphone identification},
journal = {Computer Networks},
volume = {238},
pages = {110115},
year = {2024},
issn = {1389-1286},
doi = {https://doi.org/10.1016/j.comnet.2023.110115},
url = {https://www.sciencedirect.com/science/article/pii/S1389128623005601},
author = {Xiaolin Gu and Wenjia Wu and Yusen Zhou and Aibo Song and Ming Yang and Zhen Ling and Junzhou Luo}
}

@ARTICLE{Tianya_exp_attack_2025,
  author={Zhao, Tianya and Zhang, Junqing and Mao, Shiwen and Wang, Xuyu},
  journal={IEEE Transactions on Mobile Computing}, 
  title={Explanation-Guided Backdoor Attacks Against Model-Agnostic RF Fingerprinting Systems}, 
  year={2025},
  volume={24},
  number={3},
  pages={2029-2042},
  doi={10.1109/TMC.2024.3487967}}

@INPROCEEDINGS{Zhisheng_spoofing_via_rff,
  author={Yao, Zhisheng and Wang, Yu and Gui, Guan and Otsuki, Tomoaki and Mao, Shiwen and Wang, Xianbin and Sari, Hikmet},
  booktitle={ICC 2025 - IEEE International Conference on Communications}, 
  title={A Novel Physical Spoofing Technique Using Radio Frequency Fingerprint Emulation and Model Fitting}, 
  year={2025},
  volume={},
  number={},
  pages={01-06},
  doi={10.1109/ICC52391.2025.11161944}}

@article {prisma,
	author = {Liberati, Alessandro and Altman, Douglas G and Tetzlaff, Jennifer and Mulrow, Cynthia and G{\o}tzsche, Peter C and Ioannidis, John P A and Clarke, Mike and Devereaux, P J and Kleijnen, Jos and Moher, David},
	title = {The PRISMA statement for reporting systematic reviews and meta-analyses of studies that evaluate healthcare interventions: explanation and elaboration},
	volume = {339},
	elocation-id = {b2700},
	year = {2009},
	doi = {10.1136/bmj.b2700},
	publisher = {BMJ Publishing Group Ltd},
	issn = {0959-8138},
	journal = {BMJ}
}

@inproceedings{static_sig_oligeri_2026,
author = {Oligeri, Gabriele and Sciancalepore, Savio},
title = {Beyond Static Signatures: Statistical Analysis of Radio Fingerprint Mutations},
year = {2026},
isbn = {9798400722011},
publisher = {Association for Computing Machinery},
address = {New York, NY, USA},
url = {https://doi.org/10.1145/3765613.3797445},
doi = {10.1145/3765613.3797445},
pages = {239–249},
numpages = {11},
location = {Germany},
series = {WiSec '26}
}

\begin{IEEEbiography}[{\includegraphics[width=1in,height=1.25in,clip,keepaspectratio]{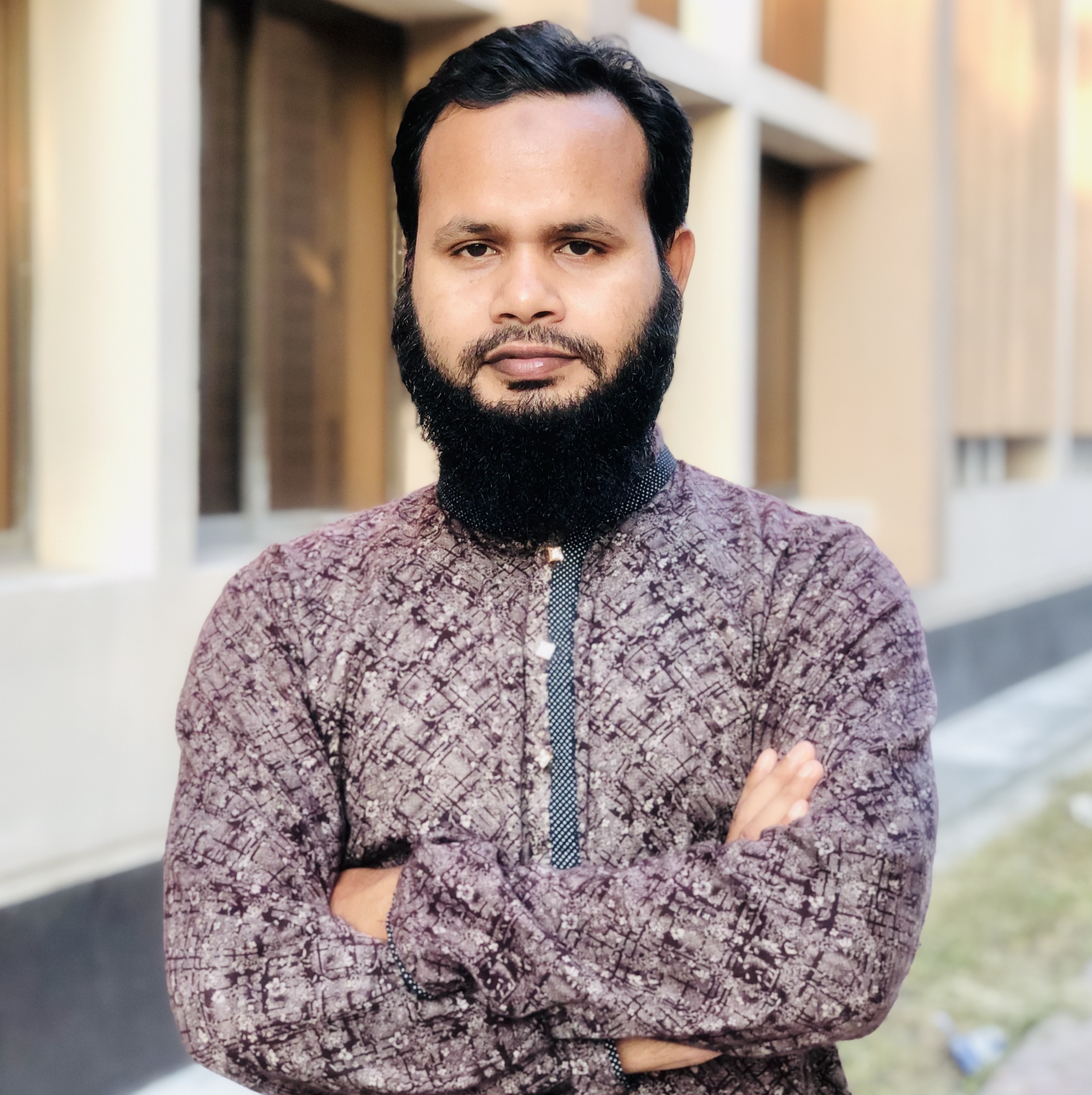}}]{Abdul Aziz} is currently pursuing his Ph.D. degree in Computer Science and Engineering program under the College of Science and Engineering at Hamad Bin Khalifa University, Doha, Qatar. He received the Bachelor's and Master's degrees in computer science and engineering from the Khulna University of Engineering \& Technology (KUET), Khulna, Bangladesh. He is also working as an Assistant Professor at KUET (on study leave now). His research interests include Radio Frequency Fingerprinting, Physical Layer Security, Deep Learning, Machine Learning, and Fuzzy Logic.
\end{IEEEbiography}

\begin{IEEEbiography}[{\includegraphics[width=1in,height=1.25in,clip,keepaspectratio]{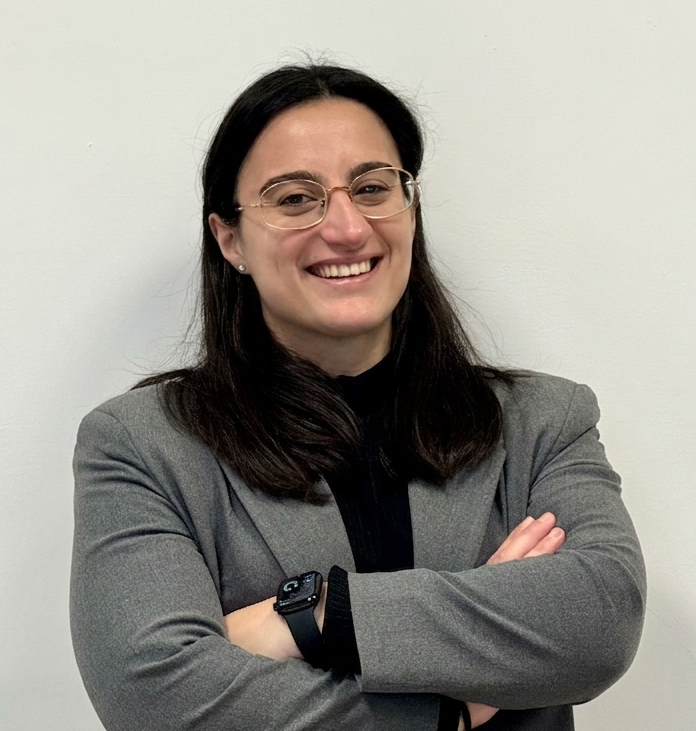}}]
{Ingrid Huso} is currently a PostDoc researcher at Hamad Bin Khalifa University, College of Science
and Engineering, Qatar. She received her Ph.D. degree in Industry 4.0 from Politecnico di Bari, Italy, in January 2025. Her main research interests include Network Security, Physical Layer Security, Privacy Enhancing Techniques, Lawful Interception, and Social Internet of Things.
\end{IEEEbiography}

\begin{IEEEbiography} [{\includegraphics[width=1in,height=1.25in,clip,keepaspectratio]{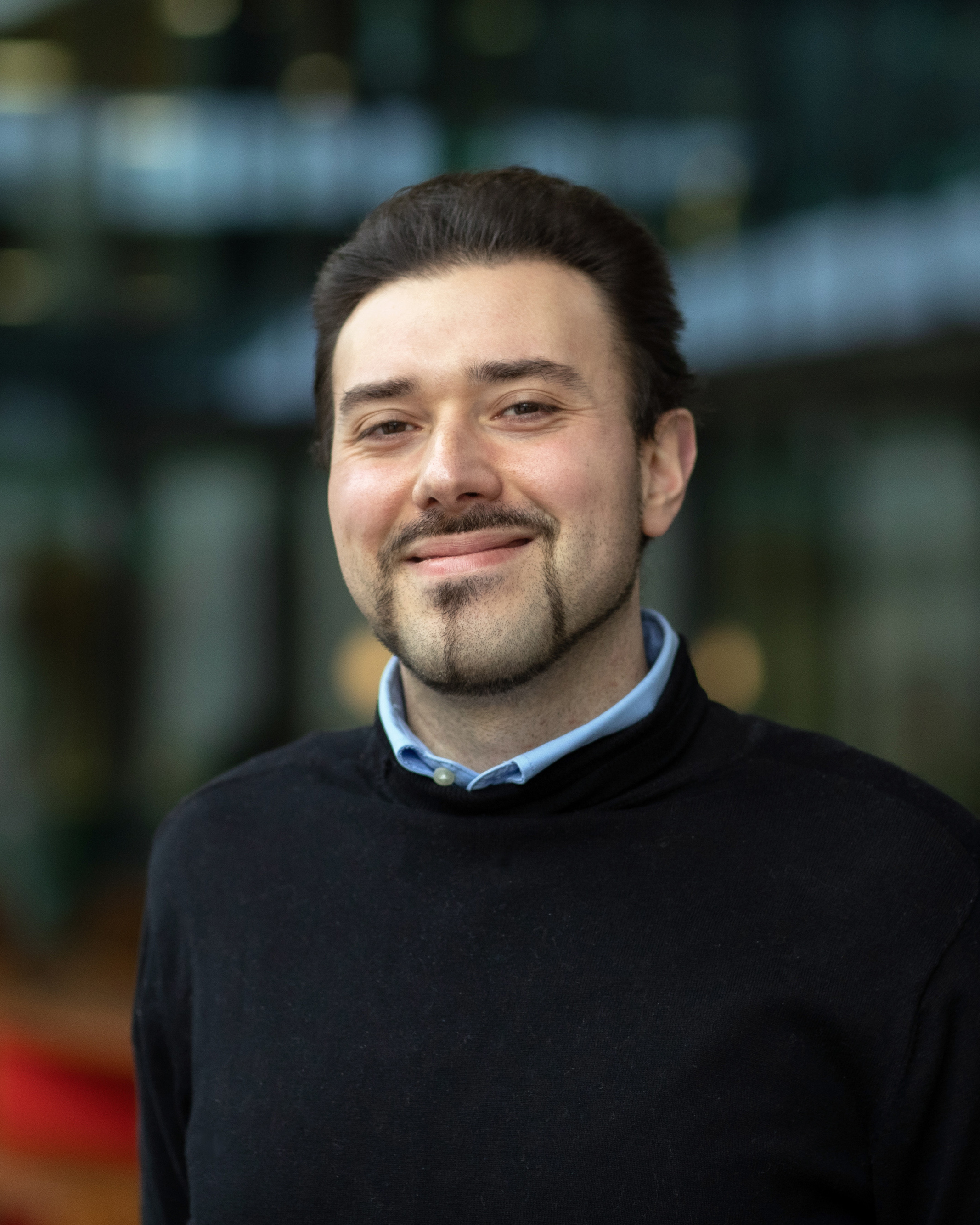}}]
{Savio Sciancalepore} is Assistant Professor in Wireless and IoT Security at TU Eindhoven, Netherlands. He received the Master’s and Ph.D. degrees from the Polytechnic University of Bari, Italy, in 2013 and 2017, respectively. Before joining TU/e in 2021, from 2017 to 2020, he was a Postdoctoral Researcher with Hamad Bin Khalifa University, College of Science and Engineering, Doha, Qatar.  
Dr. Sciancalepore received the Prestigious Award from the ERCIM Security, Trust, and Management Working Group for the Best Ph.D. Thesis in Information and Network Security in 2018. His major research interests include applied network security and privacy issues in Wireless, Mobile and Internet of Things networks, with a particular focus on the interplay between security guarantees, energy limitations, and requirements of real-world use cases. Photo by Angeline Swinkels.
\end{IEEEbiography}

\begin{IEEEbiography}[{\includegraphics[width=1in,height=1.25in, clip, keepaspectratio]{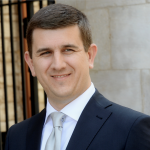}}]{Gabriele Oligeri} is an Associate Professor at Hamad bin Khalifa University, College of Science and Engineering, Qatar. He received his Ph.D. in Computer Engineering from the University of Pisa, and his research interests are in the field of signals intelligence.
\end{IEEEbiography}

\end{document}